\documentclass[onecolumn,superscriptaddress,prd,aps,preprintnumbers,amsmath,amssymb,nofootinbib]{revtex4}
\usepackage{graphicx}
\usepackage{dcolumn}
\usepackage{bm}
\usepackage{bbm}
\usepackage[pdftex]{color}
\usepackage{accents}

\newcommand{\beq}{\begin{equation}}
\newcommand{\eeq}{\end{equation}}
\newcommand{\bq}{\begin{equation}}
\newcommand{\eq}{\end{equation}}
\newcommand{\ba}{\begin{array}}
\newcommand{\ea}{\end{array}}
\newcommand{\beqa}{\begin{eqnarray}}
\newcommand{\eeqa}{\end{eqnarray}}

\def\[{\left[}
\def\]{\right]}
\def\({\left(}
\def\){\right)}

\evensidemargin=\oddsidemargin

\def\slashii#1{\setbox0=\hbox{$#1$}             
   \dimen0=\wd0                                 
   \setbox1=\hbox{\sl/} \dimen1=\wd1            
   \ifdim\dimen0>\dimen1                        
      \rlap{\hbox to \dimen0{\hfil\sl/\hfil}}   
      #1                                        
   \else                                        
      \rlap{\hbox to \dimen1{\hfil$#1$\hfil}}   
      \hbox{\sl/}                               
   \fi}                                         %
\def\slashiii#1{\setbox0=\hbox{$#1$}#1\hskip-\wd0\hbox to\wd0{\hss\sl/\/\hss}}
\def\fsl#1{\setbox0=\hbox{$#1$}           
   \dimen0=\wd0                                 
   \setbox1=\hbox{/} \dimen1=\wd1               
   \ifdim\dimen0>\dimen1                        
      \rlap{\hbox to \dimen0{\hfil/\hfil}}      
      #1                                        
   \else                                        
      \rlap{\hbox to \dimen1{\hfil$#1$\hfil}}   
      /                                         
   \fi}                                         %

\def\pslash{\not{\hbox{\kern-4pt $p$}}}
\def\qslash{\not{\hbox{\kern-4pt $q$}}}
\def\lv{\not{\hbox{\kern-4pt $L$}}}
\def\lsim{\mathrel{\raise.3ex\hbox{$<$\kern-.75em\lower1ex\hbox{$\sim$}}}}
\def\gsim{\mathrel{\raise.3ex\hbox{$>$\kern-.75em\lower1ex\hbox{$\sim$}}}}
\def\ifmath#1{\relax\ifmmode #1\else $#1$\fi}
\newcommand{\laem}{\stackrel{<}{\sim}}

\begin{document}

\preprint{MSUHEP-110927}

 \title{The Flavor Structure of the Three-Site Higgsless Model}

\author{Tomohiro Abe}
\affiliation{Institute of Modern Physics and Center for High Energy Physics,
Tsinghua University, Beijing 100084, China}

\author{R.\ Sekhar Chivukula}
\author{Elizabeth H. Simmons}
\affiliation{Department of Physics and Astronomy, Michigan State
       University, East Lansing, MI 48824, USA}

\author{Masaharu Tanabashi}  
\affiliation{Kobayashi-Maskawa Institute for the Origin of Particles and the Universe, Nagoya University, 
Nagoya 464-8602, Japan \\
and Department of Physics, Nagoya University, Nagoya 464-8602,  Japan}

\date{\today}

\begin{abstract}

We study the flavor structure of the three-site Higgsless model and evaluate the constraints on the model arising from flavor physics.  We find that current data constrain the model to exhibit only minimal flavor violation at tree level.  Moreover, at the one-loop level, by studying the leading chiral logarithmic corrections to chirality-preserving $\Delta F = 1$ and $\Delta F = 2$ processes from new physics in the model, we show that the combination of minimal flavor violation and ideal delocalization ensures that these flavor-changing effects are sufficiently small that the model remains phenomenologically viable.  
\end{abstract}

\date{\today}
 
 \maketitle


\section{Introduction}

Higgsless models \cite{Csaki:2003dt} of electroweak symmetry breaking provide effective low-energy theories of a strongly-interacting symmetry breaking sector \cite{Weinberg:1979bn,Susskind:1978ms}.  If the fermions in the model are delocalized (i.e. derive electroweak interactions from multiple gauge groups), Higgsless models can be consistent with electroweak precision
measurements \cite{Cacciapaglia:2004rb,Casalbuoni:2005rs,Cacciapaglia:2005pa,Foadi:2004ps,Foadi:2005hz,Chivukula:2005bn,SekharChivukula:2005xm} even at the loop level  \cite{SekharChivukula:2006cg,Abe:2008hb}. The three-site model \cite{SekharChivukula:2006cg} is the minimal low-energy realization of a   
Higgsless theory.\footnote{
This theory is in the same class as models of extended  electroweak gauge symmetries \cite{Casalbuoni:1985kq,Casalbuoni:1996qt}  motivated by models of hidden local symmetry \cite{Bando:1985ej,Bando:1985rf,Bando:1988ym,Bando:1988br,Harada:2003jx}. In particular the three-site model has the same
gauge structure as the ``BESS" model of \cite{Casalbuoni:1985kq}, but it is the fermion couplings and flavor structure unique to the three-site model
\cite{SekharChivukula:2006cg} that are of particular interest here.}
Its electroweak sector includes only one $SU(2)$ group beyond the usual $SU(2)\times U(1)$ of the standard model, so the gauge spectrum includes only one triplet of the extra  
vector mesons typically present in such theories; these are the mesons (denoted here by ${W'}^\pm$ and  $Z'$) that are analogous to the $\rho$ mesons of QCD. The three-site model retains sufficient complexity, however, to incorporate interesting physics issues related to fermion masses, electroweak observables, and flavor.

As discussed in \cite{SekharChivukula:2006cg} and reviewed here, the three-site model generically exhibits non-minimal flavor violation (i.e., more than the minimal flavor violation present in the standard model \cite{Chivukula:1987py, D'Ambrosio:2002ex}).  However, if one assumes that flavor-symmetry breaking enters the Lagrangian only through the delocalization parameters of the right-handed fermions ($\epsilon_{Rf}$), the three-site model then possesses only minimal flavor violation.  Moreover, if one also assumes that the (now flavor-universal) delocalization parameter $\epsilon_L$ for the left-handed fermions is set to the ``ideal" value \cite{SekharChivukula:2005xm} that correlates the fermion wavefunction with the $W$-boson wavefunction, then the tree-level electroweak phenomenology of the three-site model agrees completely with that of the standard model.  

This situation is modified once loop effects are included.  The various parameters in the effective Lagrangian, whether flavor-universal or not, will run, so the conditions of ideal delocalization and minimal flavor violation are not scale-independent. Rather, one may impose these conditions at the scale of the cutoff of the effective three-site theory -- the scale of the underlying strong dynamics -- and then compute and evaluate corrections to electroweak and flavor observables. In fact, the chiral logarithmic corrections to the flavor-universal electroweak parameters $\alpha S$ and $\alpha T$  \cite{Peskin:1990zt,Peskin:1991sw,Altarelli:1990zd,Altarelli:1991fk}  in the three-site model were computed in references \cite{Matsuzaki:2006wn,SekharChivukula:2007ic,Dawson:2007yk}; these are the one-loop contributions that dominate in the limit where the masses of the new vector mesons lie far below the cutoff of the effective theory.  Likewise, the flavor-dependent corrections to the $Z\to b\bar{b}$ branching ratio were studied in \cite{Abe:2008hb,Abe:2009ni}, and the corrections to chirality-non-preserving flavor-dependent process $b \to s\gamma$ were computed in \cite{Kurachi:2010fa}. 

This paper completes the investigation of the flavor phenomenology of the three-site model by studying the chiral logarithmic corrections to chirality-preserving flavor-changing processes.  We begin by reviewing the essential features of the model and contrasting its flavor structure with that of the standard model.  In particular, we establish the conditions under which the three-site model exhibits minimal or non-minimal flavor violation.  A brief review (with details in an Appendix) of experimental constraints on flavor-changing effects demonstrates that the tree-level Lagrangian of the three-site model is constrained to a form that, to a good approximation, has only minimal flavor violation; in the rest of the paper, we therefore assume the model exhibits only minimal flavor violation. In section IV, we calculate the corrections to all chirality preserving $\Delta F=1$ operators that arise from the new physics present in the three-site model. We show that, parametrically, the size of the new three-site corrections to $\Delta F=1$ processes are of the same order as those in the standard model -- but that the corrections numerically amount to only a few percent of the standard model contribution. Since no chirality-preserving $\Delta F=1$  neutral current standard model amplitudes are observable we conclude that, just as in the case of corrections to $Z \to b{\bar b}$, the additional three-site model chiral logarithmic contributions are not forbidden, and the three-site model remains viable. In section V, we extend our analysis to $\Delta F=2$ (meson mixing) processes.  We find that the combination of ideal delocalization and minimal flavor violation insures that the new contributions to $\Delta F = 2$ box diagrams in the three-site model are smaller than or of order two-loop corrections to these processes in the standard model Ð and hence are not phenomenologically excluded.  The final section of the paper summarizes our conclusions.

\section{The Three-Site Model}

\begin{figure}[th]
\centering
\includegraphics[width=0.40\textwidth]{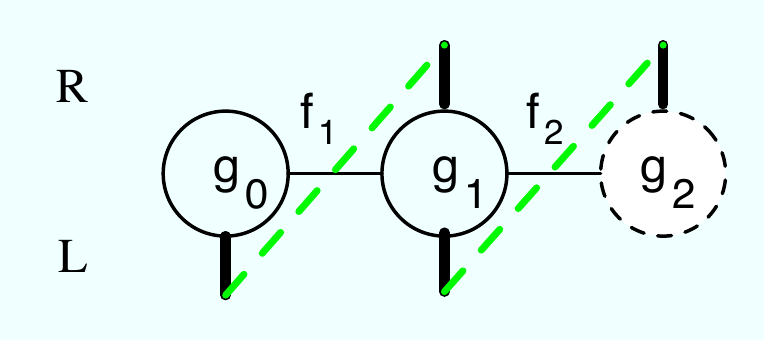}
\caption{The three-site model \protect\cite{SekharChivukula:2006cg}, illustrated using
``moose" notation \cite{Georgi:1985hf}. The solid circles represent
$SU(2)$ gauge groups, with coupling strengths $g_0$ and $g_1$, and the dashed circle is
a $U(1)$ gauge group with coupling $g_2$. The horizontal lines represent $SU(2) \times SU(2)/SU(2)$ non-linear
sigma model fields, with decay constants $f_{1,2}$,  breaking the adjacent global groups down to their diagonal sum. The left-handed fermons,
denoted by the lower vertical lines, are located at sites 0 and 1, and the right-handed fermions,
denoted by the upper vertical lines, at sites 1 and 2. The dashed green lines correspond to
Yukawa couplings, as described in the text. We will  
denote $g_0\equiv g$, $g_1\equiv \tilde{g}$, $g_2\equiv g'$ 
and take $g\,, g' \ll \tilde{g}$~.
\label{fig:one}
}
\end{figure}

The three-site $SU(2)_0 \times SU(2)_1 \times U(1)_2$ model \cite{SekharChivukula:2006cg}
is illustrated (using ``moose" notation \cite{Georgi:1985hf}) in Fig. \ref{fig:one} where, as we will see, $SU(2)_0 \times U(1)_2$
is approximately the $SU(2)_L \times U(1)_Y$ of the electroweak interactions, 
$SU(2)_1$ is a new ``hidden" gauge-symmetry \cite{Bando:1985ej,Bando:1985rf,Bando:1988br,Casalbuoni:1985kq},
and the $U(1)_2$ is embedded as the $\sigma^3$-component of an  $SU(2)_{2}$ global symmetry. We will denote the 
gauge couplings of the three groups by, $g_0\equiv g$, $g_1\equiv \tilde{g}$, and $g_2\equiv g'$
respectively.\footnote{Here $g$ and $g'$ are chosen because, as we will see, these groups are approximately
the $SU(2)_W \times U(1)_Y$ of the standard model.} The
nonlinear sigma-model and gauge-theory kinetic-energy terms in 
this model are given by
\begin{equation}
{\cal L} = \sum_{i=1,2}\,\frac{f^2_i}{4}{\rm tr}\left( D^\mu \Sigma^\dagger_i D_\mu \Sigma_i\right)
-\frac{1}{4 } (\vec{W}_0^{\mu\nu})^2-\frac{1}{4 } (\vec{W}_1^{\mu\nu})^2
-\frac{1}{4 }B^2_{\mu\nu},
\label{eq:threesite}
\end{equation}
where $\Sigma_1$ and $\Sigma_2$ are $SU(2) \times SU(2)/SU(2)$ sigma-model fields parameterized by
\begin{equation}
\Sigma_{1,2} = \exp\left(\frac{2i\pi_{1,2}}{f_{1,2}}\right)~,
\label{eq:Sigma}
\end{equation}
where $\pi_{1,2} \equiv  \pi^a_{1,2}\sigma^a/2$, and 
where $\vec{W}_{0,1}^{\mu\nu}$ and $B^{\mu\nu}$ are, respectively, 
the field-strength tensors of the $SU(2)_{0,1}$ and $U(1)_2$
gauge-groups with corresponding gauge-fields $W_{0,1}^\mu$ and $B^\mu$. 

The sigma-model fields transform as 
\begin{equation}
\Sigma_1  \to U_0\, \Sigma_1\, U_1^\dagger\label{eqn:xi1}~,\qquad\qquad
\Sigma_2 \to U_1\, \Sigma_2\, U_2^\dagger ~,
\end{equation}
under the $SU(2)_0 \times SU(2)_1 \times SU(2)_2$ global symmetries, and hence the covariant derivatives
above are given by
\begin{align}
D_\mu \Sigma_1 = \partial_\mu \Sigma_1 -i g W_{0\mu}^{a} \frac{\sigma^a}{2} \Sigma_1+i\tilde{g}W_{1\mu}^{a} \Sigma_1\frac{\sigma^a}{2}  ~,\\
D_\mu \Sigma_2 = \partial_\mu \Sigma_2 -i \tilde{g}W_{1\mu}^{a} \frac{\sigma^a}{2} \Sigma_2+ig'B_{\mu} \Sigma_2\frac{\sigma^3}{2}   ~.
\end{align}
Here $f_{1,2}$ are the $f$-constants,
the analogs of $F_\pi$ in QCD, associated with the two $SU(2) \times SU(2)/SU(2)$
nonlinear sigma-models, and they
satisfy the relation
\begin{equation}
\sqrt{2} G_F = \frac{1}{v^2} =\frac{1}{f^2_1} + \frac{1}{f^2_2}\approx
\frac{1}{(250\,{\rm GeV})^2}~.
\label{eq:GF}
\end{equation}
In \cite{SekharChivukula:2006cg}, for simplicity and to maximize the range of validity of this low-energy effective
theory, we took $f_1=f_2=\sqrt{2} v$; in this work, in order to identify the origin of various one-loop effects,
we will leave $f_{1,2}$ arbitrary, subject to the constraint in Eq. (\ref{eq:GF}) above.

In unitary gauge, $\Sigma_1=\Sigma_2 \equiv
{\cal I}$, and the non-linear sigma model kinetic terms yield vector-boson mass matrices.
We will work in the limit $g, g' \ll \tilde{g}$, or equivalently
\begin{equation}
x=g/\tilde{g} \ll 1~.
\end{equation}
We will also define an angle
$\theta$ 
\begin{equation}
\frac{g'}{g} \equiv \frac{\sin\theta}{\cos\theta}~,
\label{eq:deftheta}
\end{equation}
which will equal the usual weak mixing angle up to corrections of order $x^2$.
In the small $x$ limit, we find the charged-boson masses
\begin{equation}
M^2_W = \frac{g^2\,v^2}{4} + \ldots~,\qquad\qquad
M^2_{W'}  = \frac{\tilde{g}^2(f^2_1+f^2_2)}{4}+\ldots~,
\label{eq:Wmasses}
\end{equation}
where the mass eigenstates are of the form \cite{Casalbuoni:1985kq}
\begin{align}
W^\pm_\mu & = W^\pm_{0\mu}+\frac{x f^2_1}{f^2_1+f^2_2} W^\pm_{1\mu} + {\cal O}\left(x^2\right)~, \label{eq:Weigenstate}\\
{W'}^\pm_\mu & =-\, \frac{x f^2_1}{f^2_1+f^2_2} W^\pm_{0\mu}+ W^\pm_{1\mu} + {\cal O}\left(x^2\right)~.
\label{eq:Wpeigenstate}
\end{align}

The neutral bosons include a massless photon ($A^\mu$), which
corresponds to the eigenvector
\begin{align}
A_\mu & = \frac{e}{g} W^3_{0\mu} + \frac{e}{\tilde{g}}W^3_{1\mu} + \frac{e}{g'} B^\mu\\
& = \sin\theta  \,W^3_{0\mu} + x \sin\theta\, W^3_{1\mu} +\cos\theta\, B^\mu + {\cal O}(x^2)~,
\end{align}
where $e$ is the electromagnetic coupling
\begin{equation}
\frac{1}{e^2}=\frac{1}{g^2}+\frac{1}{\tilde{g}^2}+\frac{1}{{g'}^2}~.
\label{eqn:edef}
\end{equation}
For small $x$ we also have
\begin{equation}
g  \approx \frac{e}{\sin\theta}~,\qquad\qquad
g'  \approx \frac{e}{\cos\theta}~.
\end{equation}
The two other neutral gauge-bosons have masses
\begin{equation}
M^2_Z = \frac{e^2\,v^2}{4\sin^2\theta \cos^2\theta} + \ldots~,\qquad\qquad
M^2_{Z'}  = \frac{\tilde{g}^2 (f^2_1+f^2_2)}{4} +\ldots~,
\end{equation}
corresponding to the eigenvectors \cite{Casalbuoni:1985kq}
\begin{align}
Z_\mu & = \cos\theta\, W^{3}_{0\mu} + \frac{x\cos\theta(f^2_1-f^2_2 \tan^2\theta)}{f^2_1+f^2_2}W^3_{1\mu}
-\sin\theta\, B_\mu+{\cal O}(x^2)~,
\label{eq:Zeigenstate}\\
{Z'}_\mu & = -\,\frac{x f^2_1}{f^2_1+f^2_2} W^3_{0\mu} + W^3_{1\mu}-\frac{x\tan\theta f^2_2}{f^2_1+f^2_2}B_\mu + {\cal O}(x^2)~.
\label{eq:Zpeigenstate}
\end{align}
As described in \cite{SekharChivukula:2006cg}, working in the limit of small $x$ ($g\,, g' \ll \tilde{g}$), we get
a phenomenologically-acceptable low-energy electroweak model
if  we identify the light $W^\pm_\mu$ and $Z_\mu$ with the weak bosons, because the extra states
$W'$ and $Z'$ are much heavier than ordinary electroweak gauge bosons
($M^2_{W,Z} \ll M^2_{W', Z'}$).   In particular (after including ideal fermion delocalization \cite{SekharChivukula:2005xm})
all tree-level standard model predictions are reproduced up to corrections of order $x^4$. Note also that, in the
limit $f_1 \to \infty$ for fixed $v$, the gauge boson mass eigenstates of the three-site model reduce\footnote{While the particular
expressions for the $W$ and $Z$ mass eigenstates in Eqs. (\ref{eq:Weigenstate}) and (\ref{eq:Zeigenstate}) were calculated perturbatively for $x << 1$, the reduction
of the extended electroweak gauge to its standard model counterpart in the $f_1 \to \infty$ limit (with fixed $v$) is a more general result that 
 follows directly from the decoupling theorem \cite{Appelquist:1974tg}.} to those
of the standard model with the identification of $SU(2)_0 \times U(1)_2$ with $SU(2)_L \times U(1)_Y$.

The three-site model also incorporates the ordinary quarks and leptons, and requires the presence of additional
heavy vectorial $SU(2)_1$ fermions that  mirror the light fermions. These
heavy Dirac fermions are the analogs of the lowest Kaluza-Klein (KK) fermion modes
which would be present in an extra-dimensional theory.
The quark ``Yukawa" sector of the three-site model illustrated
in Fig. \ref{fig:one} is:
\begin{equation}
{\cal L}_{mass} = -\bar{q}^{(0)}_L \Sigma_1 \underaccent{\tilde}{\mathsf m}_1 q^{(1)}_R
-\bar{q}^{(1)}_L \underaccent{\tilde}{\mathsf M} q^{(1)}_R
-\bar{q}^{(1)}_L \Sigma_2 
\begin{pmatrix}
\underaccent{\tilde}{\mathsf m}_{2u} & 0\\
0 & \underaccent{\tilde}{\mathsf m}_{2d} 
\end{pmatrix}
\begin{pmatrix}
u^{(2)}_R \\
d^{(2)}_R
\end{pmatrix}
+h.c.~,
\label{eq:Lmass}
\end{equation}
where the quark fields $q^{(0)}_L$, $q^{(1)}_{L,R}$, $u^{(2)}_R$, and $d^{(2)}_R$ are three-component vectors
in flavor space, $\underaccent{\tilde}{\mathsf m}_1$, $\underaccent{\tilde}{\mathsf M}$, and $\underaccent{\tilde}{\mathsf m}_{2u,2d}$ are $3 \times 3$ matrices in flavor space,
and the summation over flavor and gauge indices is implicit.
The transformation properties
of the quarks under the global $SU(2)_0 \times SU(2)_1 \times SU(2)_2$ symmetries are given
by
\begin{align}
q^{(0)}_{L} & \to U_0 q^{(0)}_{L}~,\label{eq:psi0}\\
q^{(1)}_{R,L} & \to U_1 q^{(1)}_{R,L}~, \label{eqn:psi1} \\
\begin{pmatrix}
u^{(2)}_{R} \\
d^{(2)}_{R}
\end{pmatrix}
& \to U_2 
\begin{pmatrix}
u^{(2)}_{R} \\
d^{(2)}_{R}
\end{pmatrix}~.
\end{align}
The $SU(2)_0 \times SU(2)_1$ properties of the quarks follow from the assignments above; the
hypercharge properties are fixed by insuring the correct values of the electric charges, and hence
under $U(1)_2$ we require that the $q^{(0)}_L$ and $q^{(1)}_{L,R}$ fields carry charge $1/6$,
while the $u^{(2)}_R$ and $d^{(2)}_R$ carry charges $+2/3$ and $-1/3$ respectively.

We will work in the limit where the eigenvalues\footnote{More properly, the eigenvalues of $\underaccent{\tilde}{\mathsf M}\underaccent{\tilde}{\mathsf M}^\dagger$ are much greater than those of $\underaccent{\tilde}{\mathsf m}_1\underaccent{\tilde}{\mathsf m}^\dagger_1$ or  $\underaccent{\tilde}{\mathsf m}_{2u,d}\underaccent{\tilde}{\mathsf m}^\dagger_{2u,d}$.} of $\underaccent{\tilde}{\mathsf M}$ are much greater
than those for  $\underaccent{\tilde}{\mathsf m}_1$ and $\underaccent{\tilde}{\mathsf m}_{2u,d}$ and where the
heavy fermions are essentially the $q^{(1)}$ doublets
with  mass-squareds given approximately by the eigenvalues of $\underaccent{\tilde}{\mathsf M}\underaccent{\tilde}{\mathsf M}^\dagger$.  In this
limit the matrix $\epsilon_L \equiv  \underaccent{\tilde}{\mathsf m}_1 \cdot \underaccent{\tilde}{\mathsf M}^{-1}$ controls the ``delocalization" 
of the left-handed
fermions, {\it i.e.} the degree to which the light left-handed mass eigenstate
fields are admixtures of fermions at the first two sites
In \cite{SekharChivukula:2006cg}, it was assumed that $\underaccent{\tilde}{\mathsf M}$ and $\underaccent{\tilde}{\mathsf m}_1$ were
flavor-diagonal, so that $\epsilon_L$ was likewise proportional to the identity in flavor-space.  Furthermore, it was shown that the proportionality constant could be
adjusted (a process called ``ideal fermion delocalization'') to eliminate potentially dangerous tree-level
contributions to the electroweak parameter $\alpha S$
\cite{Cacciapaglia:2004rb,Casalbuoni:2005rs,Cacciapaglia:2005pa,Foadi:2004ps,Foadi:2005hz,Chivukula:2005bn,SekharChivukula:2005xm}.
In this work, we confirm that the precision electroweak and flavor data directly constrain $\epsilon_L$ to be flavor universal and close to the ideal delocalization form.  Therefore we will take $\epsilon_L$ to be flavor-universal, at tree-level in the three-site model, so that all of the flavor-breaking
is encoded in the values of Yukawa couplings to the right-handed fermions. 
As we show below, in this limit the three-site model at tree-level
has precisely the same flavor structure as the standard model: all of the tree-level
couplings of the left-handed fermions to the gauge bosons are flavor-diagonal
and equal, and flavor-changing neutral currents are suppressed 
\cite{SekharChivukula:2006cg}.  

Limits on the
$WWZ$ coupling imply that the $W'$ and $Z'$ bosons must be heavier than about 400 GeV, while
limits on the unitarity of $W_L W_L$ scattering show they must be lighter than about 1 TeV \cite{SekharChivukula:2005xm}.
On the other hand, limits on $\alpha T$ imply that the heavy fermions must have masses greater than
about 2 TeV \cite{SekharChivukula:2006cg}.

 \section{Flavor Symmetries and Structure}
  
In this section we consider the tree-level flavor structure of the three-site model. We begin with a review
of the flavor symmetries of the standard model and generalize to the three-site model.  Then, we consider the
effective Lagrangian that results from ``integrating out" the heavy fermions and analyze the tree-level
gauge-couplings.  

 \subsection{GIM Flavor Symmetries of the standard model}
 \label{subsec:flavor-SM}

Before proceeding to discuss the flavor structure of the three-site
model in detail, we first briefly review the flavor structure of the Yukawa sector
of the standard model
\begin{equation}
{\cal L}_{Yuk} = - \bar{q}_{Li} \lambda^{ij}_d \varphi d_{Rj}  - \bar{q}_{Li} \lambda^{ij}_u 
\tilde{\varphi} u_{Rj} + h.c.
\end{equation}
Here the $q_{Li}$, $u_{Rj}$, and $d_{Rj}$ fields are the three flavors of left-handed
quarks, and right-handed up- and down-type quarks respectively, $i$ and $j$ are flavor
indices, and $\lambda^{ij}_{d,u}$ are the Yukawa-coupling matrices for down-
and up-quarks. 

In the standard model, these Yukawa terms are the only interactions
that distinguish among flavors. The
gauge interactions respect an $SU(3)_L \times SU(3)_{uR} \times SU(3)_{dR}$ global
symmetry.
The Yukawa couplings $\lambda_{u,d}$ can then be treated as ``spurions", and they
can be classified by their transformation properties under these symmetries  
\cite{Chivukula:1987py}. In particular, the standard model would be invariant under
an arbitrary global flavor symmetry transformation {\it if} the Yukawa couplings transformed
as follows
\begin{equation}
\lambda_u \to L \lambda_u R^\dagger_u\qquad\qquad
\lambda_d \to L \lambda_d R^\dagger_d~,
\label{eq:yukawasymmetry}
\end{equation}
so that $\lambda_{u,d}$ transformed, respectively, as elements
of the $(3,\bar{3},1)$ and $(3,1,\bar{3})$ representations under $SU(3)_L \times SU(3)_{uR} \times SU(3)_{dR}$.

The $SU(3)_L \times SU(3)_{uR} \times SU(3)_{dR}$ symmetries are sufficient
to diagonalize {\it either} $\lambda_u$ or $\lambda_d$. Therefore, 
there can be no tree-level flavor-changing neutral currents: we can always
choose to work in a basis in which $\lambda_d$ (for example) is diagonal, and
in this basis the $Z$-boson will not connect quarks from different generations.
In other words, the $SU(3)_L \times SU(3)_{uR} \times SU(3)_{dR}$ global symmetry 
underlies the GIM mechanism \cite{Glashow:1970gm}. The same, of course,
is {\it not} true of the charged weak currents: the mismatch in the $L$ transformations
required to diagonalize the $\lambda_u$ and $\lambda_d$ couplings results
in the CKM matrix \cite{Cabibbo:1963yz,Kobayashi:1973fv}.
In addition, to the extent that the $\lambda_{u,d}$ are {\it small} parameters,
flavor-violating effects are suppressed by various powers of these couplings.
The flavor transformation properties of the amplitudes that give rise to
these flavor-violating effects can be used to understand the structure 
and order of magnitude of the leading standard model 
contributions.\footnote{One subtlety in this type of reasoning is worth emphasizing: sometimes,
in cases that correspond to ``long-distance" effects, some of the dependence on the quark
masses is non-analytic. This explains, for example, why the ``box diagram" contributions
to $\Delta S = 2$ processes in the standard model appear to be suppressed only
by {\it two} powers of quark masses instead of the four powers one would expect
on the basis of flavor symmetries -- two powers of quark mass appear in the denominator
after loop integration, canceling   two in the numerator that are there due to the flavor and chiral structure.}

The same reasoning can be extended beyond the standard model as well: by 
classifying the flavor-transformation properties of the new interactions, one
can understand the structure and order of magnitude of flavor-violating 
processes in these new theories.  From a symmetry point of view, the {\it minimal} amount of flavor violation in any theory is that 
which exists in the standard model \cite{Chivukula:1987py}.  In particular, the quark sector of any theory must include ``spurions" 
that transform as $(3,\bar{3},1)$ and $(3,1,\bar{3})$ under 
$SU(3)_L \times SU(3)_{uR} \times SU(3)_{dR}$ to account for the observed
quark masses. This idea has been termed ``Minimal Flavor Violation" \cite{D'Ambrosio:2002ex}.
Any new interactions in the model should, otherwise, be as
 flavor-symmetric as possible in order to avoid generating large flavor-changing neutral currents.

\subsection{Flavor Structure of the Three-Site Model at Tree Level}
\label{subsec:flavor-three-site-tree}

We now examine the flavor structure of the three-site model.  We begin
by defining the global symmetry group $SU(3)_L \times SU(3)_{LD} \times SU(3)_{RD}
\times SU(3)_{uR} \times SU(3)_{dR}$ under which the fields transform as:
\begin{align}
q^{(0)}_L & \to L \cdot q^{(0)}_L \label{eq:symm}\\
q^{(1)}_L & \to L_D \cdot q.^{(1)}_L\nonumber\\
q^{(1)}_R & \to R_D \cdot q^{(1)}_R\nonumber\\
u^{(2)}_R & \to R_u \cdot u^{(2)}_R \nonumber\\
d^{(2)}_R & \to R_d \cdot d^{(2)}_R~,\nonumber
\end{align}
where $L$, $L_D$, $R_D$, $R_u$, and $R_d$ are arbitrary elements of
$SU(3)_L$, $SU(3)_{LD}$, $SU(3)_{RD}$, 
$SU(3)_{uR}$, and $SU(3)_{dR}$ respectively. These symmetries are broken by the interactions in Eq. (\ref{eq:Lmass}), and the
various masses are ``spurions" -- in particular, the theory would be invariant
under $SU(3)_L \times SU(3)_{LD} \times SU(3)_{RD}
\times SU(3)_{uR} \times SU(3)_{dR}$ transformations if the mass-parameters were
simultaneously changed as follows:
\begin{align}
\underaccent{\tilde}{\mathsf m}_1 & \to L \cdot \underaccent{\tilde}{\mathsf m}_1 \cdot R^\dagger_D \label{eq:spurion}\\
\underaccent{\tilde}{\mathsf M} & \to L_D \cdot \underaccent{\tilde}{\mathsf M} \cdot R^\dagger_D \nonumber\\
\underaccent{\tilde}{\mathsf m}_{2u} & \to L_D \cdot \underaccent{\tilde}{\mathsf m}_{2u} \cdot R^\dagger_u \nonumber\\
\underaccent{\tilde}{\mathsf m}_{2d} & \to L_D \cdot \underaccent{\tilde}{\mathsf m}_{2d} \cdot R^\dagger_d~.\nonumber
\end{align}

Of course the mass matrices in Eq. (\ref{eq:Lmass}) 
are fixed, and do not transform -- so their presence
breaks the flavor symmetries. In general, without any further assumptions about
the structure of these masses, one could go to a basis where $\underaccent{\tilde}{\mathsf m}_1$
and either $\underaccent{\tilde}{\mathsf m}_{2u}$ or $\underaccent{\tilde}{\mathsf m}_{2d}$ are diagonal -- but one would
not have freedom to diagonalize either $\underaccent{\tilde}{\mathsf m}_2$  or $\underaccent{\tilde}{\mathsf M}$. This shows, as expected,
that without further assumptions about the masses the three-site model has non-minimal
flavor violation.

Combining the left- and right-handed quarks into twelve-component vectors (suppressing flavor indices)
\begin{equation}
{\cal Q}_L = \begin{pmatrix}
q^{(0)}_L= \begin{pmatrix}
u^{(0)}_L \\
d^{(0)}_L
\end{pmatrix}\\
\\
q^{(1)}_L=\begin{pmatrix}
u^{(1)}_L \\
d^{(1)}_L
\end{pmatrix}
\end{pmatrix}
\hspace{1.5 cm}
{\cal Q}_R = \begin{pmatrix}
q^{(2)}_R =\begin{pmatrix}
u^{(2)}_R \\
d^{(2)}_R
\end{pmatrix}
 \\
 \\
q^{(1)}_R =\begin{pmatrix}
u^{(1)}_R \\
d^{(1)}_R
\end{pmatrix}
\end{pmatrix}~,
\end{equation}
the $12\times 12$ mass matrix for the quark sector may be written (each block is $6 \times 6$)
\begin{equation}
{\cal M} = 
\begin{pmatrix}
0 & & \Sigma_1 \otimes \underaccent{\tilde}{\mathsf m}_1 \\
\\
\Sigma_2 \otimes \begin{pmatrix}
\underaccent{\tilde}{\mathsf m}_{2u} & 0\\
0 & \underaccent{\tilde}{\mathsf m}_{2d} 
\end{pmatrix}
 & & {\cal I}_{2\times 2}\otimes \, \underaccent{\tilde}{\mathsf M}
 \end{pmatrix}~,
 \label{eq:quarkmasses}
\end{equation}
where we include the factors of $\Sigma_{1,2}$ so as to maintain the $SU(2)_0 \times SU(2)_1 \times SU(2)_2$ global
symmetry and, hence, an $SU(2)_0 \times SU(2)_1 \times U(1)_2$ gauge invariance.
In the limit in which the eigenvalues of $\underaccent{\tilde}{\mathsf M}$ are larger than those of
$\underaccent{\tilde}{\mathsf m}_1$, $\underaccent{\tilde}{\mathsf m}_{2u}$, or $\underaccent{\tilde}{\mathsf m}_{2d}$, this matrix has the usual
``seesaw" form. It is convenient to define the $3 \times 3$ flavor-space matrices
\begin{align}
\epsilon_L = \underaccent{\tilde}{\mathsf m}_1 \cdot \underaccent{\tilde}{\mathsf M}^{-1} & \to L \cdot \epsilon_L \cdot L^\dagger_D~, \label{eq:epsilonL}\\
\epsilon_{Ru} = \underaccent{\tilde}{\mathsf m}^\dagger_{2u} \cdot \left(\underaccent{\tilde}{\mathsf M}^\dagger\right)^{-1} & \to R_u \cdot \epsilon_{Ru} \cdot R^\dagger_D~, \\
\epsilon_{Rd} = \underaccent{\tilde}{\mathsf m}^\dagger_{2d} \cdot \left(\underaccent{\tilde}{\mathsf M}^\dagger\right)^{-1} & \to R_d \cdot \epsilon_{Rd} \cdot R^\dagger_D ~, \label{eq:epsilonRd}
\end{align}
which, from Eq. (\ref{eq:symm}), have the flavor transformation properties indicated. The elements of these matrices are, in the
seesaw limit, small quantities. Diagonalizing ${\cal M}{\cal M}^\dagger$ and ${\cal M}^\dagger {\cal M}$,
we  find the light and heavy mass eigenstate fields
$q$ and $Q$, whose components are approximately related to the gauge-eigenstate fields by\footnote{The sign
convention of the  fields was chosen to agree with \protect\cite{SekharChivukula:2006cg}.}
\begin{align}
q^{(0)}_L & \simeq -q_{L} + \Sigma_1 \epsilon_L Q_{L} \label{eq:light}\\
q^{(1)}_L & \simeq Q_{L} + \epsilon^\dagger _L \Sigma^\dagger_1 q_{L}~,
\end{align}
and
\begin{align}
q^{(2)}_R & \simeq q_{R} + \begin{pmatrix}
\epsilon_{Ru} & 0\\
0 & \epsilon_{Rd}
\end{pmatrix} 
\Sigma^\dagger_2 Q_{R}\\
q^{(1)}_R & \simeq Q_{R} - \Sigma_2 \begin{pmatrix}
\epsilon^\dagger_{Ru} & 0\\
0 & \epsilon^\dagger_{Rd}
\end{pmatrix}
q_{R}~.
\label{eq:heavy}
\end{align}
Here, for convenience, we have chosen fields $q_{L}$, $q_{R}$, and $Q_{L,R}$ to transform under
the $SU(2)_0$, $SU(2)_2$, and $SU(2)_1$ global symmetry groups respectively.

To investigate flavor phenomenology in the three-site model we may ``integrate out" the heavy Dirac
Fermions $Q$ at tree-level. Keeping terms with two factors of the small $\epsilon$ matrices, this corresponds to inserting Eqs. (\ref{eq:light}) - (\ref{eq:heavy})
into the fermion three-site model Lagrangian, and setting the heavy fields $Q \equiv 0$.
Doing so, we obtain:
\begin{align}
{\cal L}_{eff} & = \bar{q}_{L} i \fsl{D} q_{L}+\bar{u}_{R} i \fsl{D} u_{R} + \bar{d}_{R} i \fsl{D} d_{R} 
-\left[ 
\bar{q}_{L} \epsilon_L  \Sigma_1 \Sigma_2 \underaccent{\tilde}{\mathsf M} 
\begin{pmatrix}
\epsilon^\dagger_{Ru} & 0\\
0 & \epsilon^\dagger_{Rd}
\end{pmatrix}
\begin{pmatrix}
u_{R}\\
d_{R}
\end{pmatrix}
+h.c.\right] \label{eq:Efflag}\\
& + \bar{q}_{L}\epsilon_L \left[\gamma^\mu \Sigma_1(i D_\mu \Sigma^\dagger_1)\right]
 \epsilon^\dagger_L q_{L}
 + \bar{q}_{R} 
\begin{pmatrix}
\epsilon_{Ru} & 0\\
0 & \epsilon_{Rd}
\end{pmatrix}\left[\gamma^\mu \Sigma^\dagger_2 (iD_\mu \Sigma_2)\right]
\begin{pmatrix}
\epsilon^\dagger_{Ru} & 0\\
0 & \epsilon^\dagger_{Rd}
\end{pmatrix}
q_{R}~.\nonumber
\end{align}
Here we have neglected terms that result purely in wavefunction renormalization of the
fermion fields, and terms of ${\cal O}(\epsilon^3)$. 
An important check on this result is that all of the terms in Eq. (\ref{eq:Efflag})
 are invariant under an arbitrary
$SU(3)_L \times SU(3)_{LD} \times SU(3)_{RD} \times SU(3)_{uR} \times SU(3)_{dR}$
transformation, Eq. (\ref{eq:symm}), {\it combined with} the
spurion parameter change in Eq. (\ref{eq:spurion}).
We emphasize that Eq. (\ref{eq:Efflag}) is entirely basis independent -- and therefore any
results derived from it are parameterization and phase independent as well.

The last term on the first line of Eq. (\ref{eq:Efflag}) yields the up- and down-quark masses
\begin{align}
{\cal M}_u = \epsilon_L \underaccent{\tilde}{\mathsf M} \epsilon^\dagger_{Ru} & \to L\cdot {\cal M}_u \cdot R^\dagger_u \\
 {\cal M}_d = \epsilon_L \underaccent{\tilde}{\mathsf M} \epsilon^\dagger_{Ru} & \to L\cdot {\cal M}_u \cdot R^\dagger_d~,
 \end{align}
 which transform precisely as the Yukawa couplings in the standard model, Eq. (\ref{eq:yukawasymmetry}). 
Without loss of generality, we may write the most general quark mass matrices as
\begin{equation}
{\cal M}_u =  \Lambda_u \Delta_u P^\dagger_u~,
\end{equation}
for up-quarks, and 
\begin{equation}
{\cal M}_d =\Lambda_d \Delta_d P^\dagger_d~,
\end{equation}
for down-quarks. Here $\Delta_{u,d}$ are the diagonal up- and down-quark mass
matrices, with all masses positive, and $\Lambda_{u,d}$ and $P_{u,d}$ are arbitrary
unitary matrices.\footnote{Here and throughout this note we assume the
freedom to make arbitrary phase redefinitions of the quark fields. In principle, due to 
the axial anomaly, these redefinitions will be accompanied by a change in the QCD
$\bar{\theta}$ parameter.}
Just as in the standard model the
$SU(3)_L \times SU(3)_{uR} \times SU(3)_{dR}$ subgroup of the three-site flavor
symmetry group is sufficient to diagonalize either the mass matrix of the up- or down-type quarks, 
 but not both simultaneously. 
In a basis in which the down-quark masses are diagonal, from Eq. (\ref{eq:spurion}), we have
\begin{align}
{\cal M}_d & = \Delta_d \label{eq:diagonal-down-masses}\\
{\cal M}_u & = (\Lambda^\dagger_d \Lambda_u) \Delta_u \equiv V^\dagger_{CKM} \Delta_u~,
\label{eq:diagonal-up-masses}
\end{align}
where $V_{CKM}$ is the usual quark-mixing matrix. Note also that the field $\Sigma_1 \Sigma_2$ in the
last term of the first line of Eq. (\ref{eq:Efflag})
contains precisely the unphysical Goldstone boson $\pi_W$ corresponding
to the light $W$ gauge-boson.

 The presence of the additional terms in the second line of Eq. (\ref{eq:Efflag}), involving $\epsilon_L$, $\epsilon_{Ru}$, and $\epsilon_{Rd}$, implies
 that the three-site model generically includes non-minimal flavor violation. To miminize the amount of flavor violation in the model, as discussed
 in \cite{SekharChivukula:2006cg}, we will assume\footnote{This situation is similar to the assumed flavor-universality of 
 soft SUSY breaking masses in supersymmetric extensions of the standard model.}  that both $\underaccent{\tilde}{\mathsf m}_1$ and $\underaccent{\tilde}{\mathsf M}$ are {\it flavor-universal}, and proportional to the identity matrix
 \begin{align}
\underaccent{\tilde}{\mathsf M} \equiv M \cdot {\cal I}_{3 \times 3} \label{eq:defM}~,\\
\underaccent{\tilde}{\mathsf m}_1 \equiv m_1 \cdot {\cal I}_{3 \times 3}~,
\end{align}
except where explicitly stated otherwise. If $\underaccent{\tilde}{\mathsf m}_1, \underaccent{\tilde}{\mathsf M} \propto {\cal I}_{3\times 3}$ then, from Eqs. (\ref{eq:epsilonL} -- \ref{eq:epsilonRd}) and in the basis in which ${\cal M}_d$ is diagonal, 
 \begin{align}
 \epsilon_L & \propto {\cal I}~,\label{eq:epsilonL-diagonal}\\
 \epsilon_{Ru} &  \propto V^\dagger_{CKM} \Delta_u~,\\
 \epsilon_{Rd} & \propto \Delta_d~. \label{eq:epsilonR-diagonal}
 \end{align}
 Here we see explicitly that all flavor-violation is precisely of a form determined by the quark-mass matrices, as
 expected in a minimally flavor-violating theory.  This assumption is also directly supported by constraints from precision electroweak data, and data on flavor violation in the charged-lepton and quark sectors, as we will summarize in section \ref{subsec:expt-epsilon-identity}  and explain in the Appendix. 
  
 \subsection{Gauge-Boson Couplings at Tree-Level}
 \label{subsec:gauge-coupl-tree}
 
The light quark fields $q_L,\, u_R$ and $d_R$ in the effective Lagrangian of Eq. (\ref{eq:Efflag}) couple only
to the $SU(2)_0\times U(1)_2$ gauge-eigenstate fields
\begin{align}
D_\mu q_L = \left[ \partial_\mu - igW^a_{0\mu}\frac{\sigma^a}{2} - i g' \frac{B_\mu}{6}\right] q_L~,\label{eq:qLkinetic}\\
D_\mu \begin{pmatrix}
u_R \\
d_R
\end{pmatrix} 
= \left[ \partial_\mu - i g' B_\mu \begin{pmatrix}
\frac{2}{3} & 0\\
0 & -\frac{1}{3}
\end{pmatrix}
\right]
\begin{pmatrix}
u_R \\
d_R
\end{pmatrix} ~.
\end{align}
Using Eqs. (\ref{eq:Weigenstate} -- \ref{eq:Wpeigenstate}) and
(\ref{eq:Zeigenstate} -- \ref{eq:Zpeigenstate}), the  fermion kinetic energy terms give
the conventional couplings of the light $W$ and $Z$ bosons to the quarks. From Eqs.
(\ref{eq:diagonal-down-masses} -- \ref{eq:diagonal-up-masses}), we see that these
interactions have the same flavor structure as in the standard model.
The fermion kinetic energy terms also give rise to couplings of the light quarks to 
the heavy gauge bosons
\begin{equation}
-\,\frac{g}{\sqrt{2}} \frac{x f^2_1}{f^2_1+f^2_2} {W'}^\pm_\mu \sigma^\pm -
\left(
 \frac{g\, x f^2_1}{f^2_1+f^2_2} \frac{\sigma^3}{2} + \frac{g' x f^2_2}{f^2_1+f^2_2} Y\right) 
{Z'}_\mu+ {\cal O}(x^2)~,
\label{eq:newheavycouplings}
\end{equation}
where the $\sigma^{\pm,3}$ and $Y$ encode the $SU(2) \times U(1)_Y$ quantum numbers of the quark. As expected for minimal flavor violation, there are no tree-level flavor-changing neutral currents and the
charged-current flavor structure is determined by the CKM mixing matrix.

In addition,  the terms on the second line Eq. (\ref{eq:Efflag}) give rise to additional tree-level
couplings to the gauge-bosons.  In unitary gauge, 
 we see that these terms give rise to terms involving the neutral and charged gauge-bosons
 \begin{align}
\epsilon_L \Sigma_1 (iD_\mu \Sigma^\dagger_1) \epsilon^\dagger_L \to \left(-g W^a_{0\mu}+\tilde{g}W^a_{1\mu}\right) \frac{\sigma^a}{2} \epsilon_L \epsilon^\dagger_L& =
\begin{cases}
\left(-\, \frac{g}{\sqrt{2}} \frac{f^2_2}{f^2_1+f^2_2} W^\pm_\mu + \frac{\tilde{g}}{\sqrt{2}} {W'}^\pm_\mu\right) \sigma^\pm \epsilon_L \epsilon^\dagger_L~, & a=\pm \\
\left(-\, g \frac{f^2_2}{\cos\theta(f^2_1+f^2_2)} Z_\mu + \tilde{g} {Z'}_\mu\right)\frac{\sigma^3}{2} \epsilon_L \epsilon^\dagger_L~,
& a=3  \end{cases}~,
\label{eq:newleft}\\
\epsilon_R \Sigma^\dagger_2 (i D_\mu \Sigma_2)\epsilon^\dagger_R \to \epsilon_R \left(\tilde{g}W^a_{1\mu} \frac{\sigma^a}{2}- g' B_\mu\frac{\sigma^3}{2}\right)\epsilon^\dagger_R & =
\begin{cases}
\left(\frac{g}{\sqrt{2}} \frac{f^2_1}{f^2_1+f^2_2} W^\pm_\mu + \frac{\tilde{g}}{\sqrt{2}} {W'}^\pm_\mu\right)\epsilon_R\sigma^\pm \epsilon^\dagger_R~, & a=\pm \\
\left(g \frac{f^2_1}{\cos\theta(f^2_1+f^2_2)} Z_\mu + \tilde{g} {Z'}_\mu\right) \epsilon_R \frac{\sigma^3}{2} \epsilon^\dagger_R~, & a=3
\end{cases}~,
\label{eq:newright}
\end{align}
where, for convenience, we have defined
\begin{equation}
\epsilon_R \equiv \begin{pmatrix}
\epsilon_{Ru} & 0 \\
 0 & \epsilon_{Rd}
\end{pmatrix}~.
\end{equation}

Using Eqs. (\ref{eq:epsilonL-diagonal} -- \ref{eq:epsilonR-diagonal}) we again see that there are no flavor-changing
neutral currents at tree-level, and that the strengths of charged-current processes are proportional to the CKM matrix. Comparing
Eqs. (\ref{eq:newleft}) and (\ref{eq:qLkinetic}), we see that the light-fermion portions of
the $SU(2)$ currents to which the $W$ and $Z$ bosons
couple are
\begin{equation}
j^{a\mu}_L \supset \bar{q}_L \gamma^\mu \frac{\sigma^a}{2} \left(1 -\frac{ \epsilon_L \epsilon^\dagger_L\, f^2_2}{f^2_1 + f^2_2}\right) q_L~,
\label{eq:jLcurrent}
\end{equation}
consistent with equation (27) of \cite{Abe:2009ni}.\footnote{Note here, again, that in the limit $f_1 \to \infty$ and with $v$ fixed the
three-site model reduces to the standard model -- in this case for the light fermion couplings as well.}

Combining the terms in Eq. (\ref{eq:newleft}) with those in Eq. (\ref{eq:newheavycouplings}), we see that
the $W'$ couplings to light fermions are proportional to
\begin{equation}
\tilde{g} \epsilon_L \epsilon^\dagger_L - \frac{g\, x\, f^2_1}{f^2_1+f^2_2}~.
\end{equation}
Hence, if $\epsilon_L$ is flavor-universal and satisfies
\begin{align}
\epsilon_L \epsilon^\dagger_L &  = \frac{x^2 f^2_1}{f^2_1+f^2_2} \cdot {\cal I} + {\cal O}(x^4) 
\label{eq:ideal-delocalization}\\
& = \frac{f^2_1}{v^2}\,\frac{M^2_W}{M^2_{W'}} \cdot {\cal I} + {\cal O}(x^4)~,\label{eq:ideal-def}
\end{align}
the couplings of the light fermions to the $W'_\mu$  {\it vanish}, along with the $T^3$ coupling of the $Z'_\mu$.
Defining
\begin{equation}
(\varepsilon^{ideal}_L)^2 = \frac{x^2 f^2_1}{f^2_1+f^2_2}= \frac{f^2_1}{v^2}\,\frac{M^2_W}{M^2_{W'}}~,
\end{equation}
we see that  $\epsilon^\dagger_L \epsilon_L = (\varepsilon^{ideal}_L)^2 \cdot {\cal I}$ 
is equivalent to the ``ideal fermion delocalization" condition of ref. \cite{SekharChivukula:2005xm}. As we
demonstrate in the Appendix, this amount of delocalization
insures the equality of the tree-level three-site model couplings to those of
the standard model, up to corrections of order $x^4$ \cite{SekharChivukula:2005xm} and the absence of large tree-level corrections to precision electroweak
measurements \cite{Cacciapaglia:2004rb,Casalbuoni:2005rs,Cacciapaglia:2005pa,Foadi:2004ps,Foadi:2005hz,Chivukula:2005bn,SekharChivukula:2005xm,SekharChivukula:2006cg}. The terms in equation (\ref{eq:newright}) can, however, yield small, and potentially flavor-dependent,
{\it right-handed} \cite{Foadi:2005hz,SekharChivukula:2006cg} $W$-couplings proportional to the product of
the masses of the quarks involved.

\subsection{Experimental constraints on  $\epsilon_L$}
\label{subsec:expt-epsilon-identity}

As stated earlier, assuming $\epsilon_L$ is proportional to the identity matrix minimizes the amount of flavor violation in the model and assuming the proportionality constant comes from equation  (\ref{eq:ideal-def}) minimizes the size of precision electroweak corrections.  Here, we note that precision electroweak measurements and bounds on flavor-violation in the charged-lepton and quark sectors specifically constrain $\epsilon_L$ to take this  same ``ideal delocalization'' form.

Starting with the quark sector, we adopt the basis in which the down-quark mass matrix is diagonal.  Then the elements of $(\epsilon_L \epsilon^\dagger_L) \equiv \eta$ potentially induce flavor-dependent $Z$ and $Z'$ couplings to quarks.    In other words, we are interested in the degree to which experiment allows this matrix to depart from the form in equation (\ref{eq:ideal-def}), where each diagonal element has the value $(\varepsilon_L^{ideal})^2 = \frac{f_1^2}{v^2} \frac{M_W^2}{M_{W'}^2}$ and the off-diagonal elements simply vanish.  As detailed in the Appendix, data on flavor-changing neutral currents in the B-meson, Kaon, and D-meson systems and $Z$-pole measurements of the rate at which the $Z$ decays to heavy quarks, as opposed to all hadrons, require at 90\%CL that (here we bound the absolute value of each matrix element)
\begin{equation}
\vert \eta - (\varepsilon^{ideal}_L)^2\cdot {\cal I} \vert  \laem 
(\varepsilon_L^{ideal})^2 \left(\frac{M_{W'}}{400\ {\rm GeV}}\right)^2 \left(\begin{array}{ccc}0.30 &  0.0060 \frac{\sqrt{2}v}{f_1} &  0.0285 \frac{\sqrt{2}v}{f_1} \\ 
0.0060 \frac{\sqrt{2}v}{f_1} & 0.30 & 0.202 \frac{\sqrt{2}v}{f_1} \\ 0.0285 \frac{\sqrt{2}v}{f_1}&  0.202 \frac{\sqrt{2}v}{f_1} & 0.09\end{array}\right)\,,
\label{eq:defineeta}
\end{equation}
subject to the further constraint that the first two diagonal elements must be nearly identical
\begin{equation}
|\eta_{11}-\eta_{22}| \le 2.61 \times 10^{-3}   \left(\frac{f_1}{\sqrt{2}v}\right) = 0.0323 \, (\varepsilon^{ideal}_L)^2\left(\frac{M_{W'}}{400\ {\rm GeV}}\right)^2\left(\frac{\sqrt{2}v}{f_1}\right)\, .
\end{equation}
In other words, experiment essentially constrains $\eta$ to be of the form shown in (\ref{eq:ideal-def}).

Analogously, in the charged-lepton sector, we adopt the basis in which the charged lepton mass matrix is diagonal and ignore neutrino masses.  Then the elements of $(\epsilon_L \epsilon^\dagger_L)_{lepton} \equiv \eta_\ell$ potentially induce flavor-dependent $Z$ and $Z'$ couplings to the charged leptons.  Again,  we are interested in the degree to which experiment allows this matrix to depart from the form in equation (\ref{eq:ideal-def}).  LEPEWWG bounds on the $Z$ boson's decay rates into charged leptons and on $Z$-pole leptonic charge asymmetries, as well as searches for the flavor-violating decays $\mu \to 3 e$, $\tau \to e \mu\mu$ and $\tau \to \mu e e$, combine to require at 90\%CL that
(again, we bound the absolute value of each matrix element)
\begin{equation}
|\eta_\ell - (\varepsilon^{ideal}_L)^2\cdot {\cal I}| \laem (\varepsilon_L^{ideal})^2 \left(\frac{M_{W'}}{400\ {\rm GeV}}\right)^2 
\left(
\begin{array}{ccc}
0.036 & 0.00013 &  0.034 \\ 
0.00013 & 0.075 &  0.036 \\ 
 0.034 &  0.036 & 0.12\end{array}\right)
\end{equation}
so that the matrix must have the form of (\ref{eq:ideal-def}).  Again, details are given in the Appendix.

\section{$\Delta F=1$ Processes at One Loop}

If $\underaccent{\tilde}{\mathsf m}_1$ and $\underaccent{\tilde}{\mathsf M}$ are assumed to be flavor-diagonal and 
the ratio $\epsilon_L$ is chosen to yield ideal delocalization, then tree-level three-site model electroweak
phenomenology agrees with the standard model. The situation is modified at the loop
level, however. The effective Lagrangian parameters $\underaccent{\tilde}{\mathsf m}_1$, $\underaccent{\tilde}{\mathsf M}$, and $\underaccent{\tilde}{\mathsf m}_{2u,2d}$ run in
the usual way, and therefore the conditions of ideal delocalization and minimal flavor violation are not scale-independent.
Rather, we may impose these conditions at the scale of the cutoff $\Lambda$ of the effective three-site theory and then compute the chiral-logarithmic corrections to observables at accessible energy scales.

In this section, we consider the three-site corrections
to all chirality preserving $\Delta F=1$ operators, and review the results of \cite{Kurachi:2010fa}
on the chirality non-preserving process $b \to s \gamma$. We show that, parametrically, the sizes of the new
three-site corrections to $\Delta F=1$ processes are of the same order as those in the standard model -- but
that the corrections numerically amount to only a few \% of the standard model contribution. We conclude
that, just as in the case of corrections to $Z \to b{\bar b}$, the additional three-site model chiral logarithmic contributions
are not forbidden, and the three-site model is consistent with data. In the next section we extend our analysis
to $\Delta F=2$ (meson mixing) processes.

\subsection{$Z \to \bar{f} f'$}

\begin{figure}[tb]
\includegraphics[width=5cm,height=5cm]{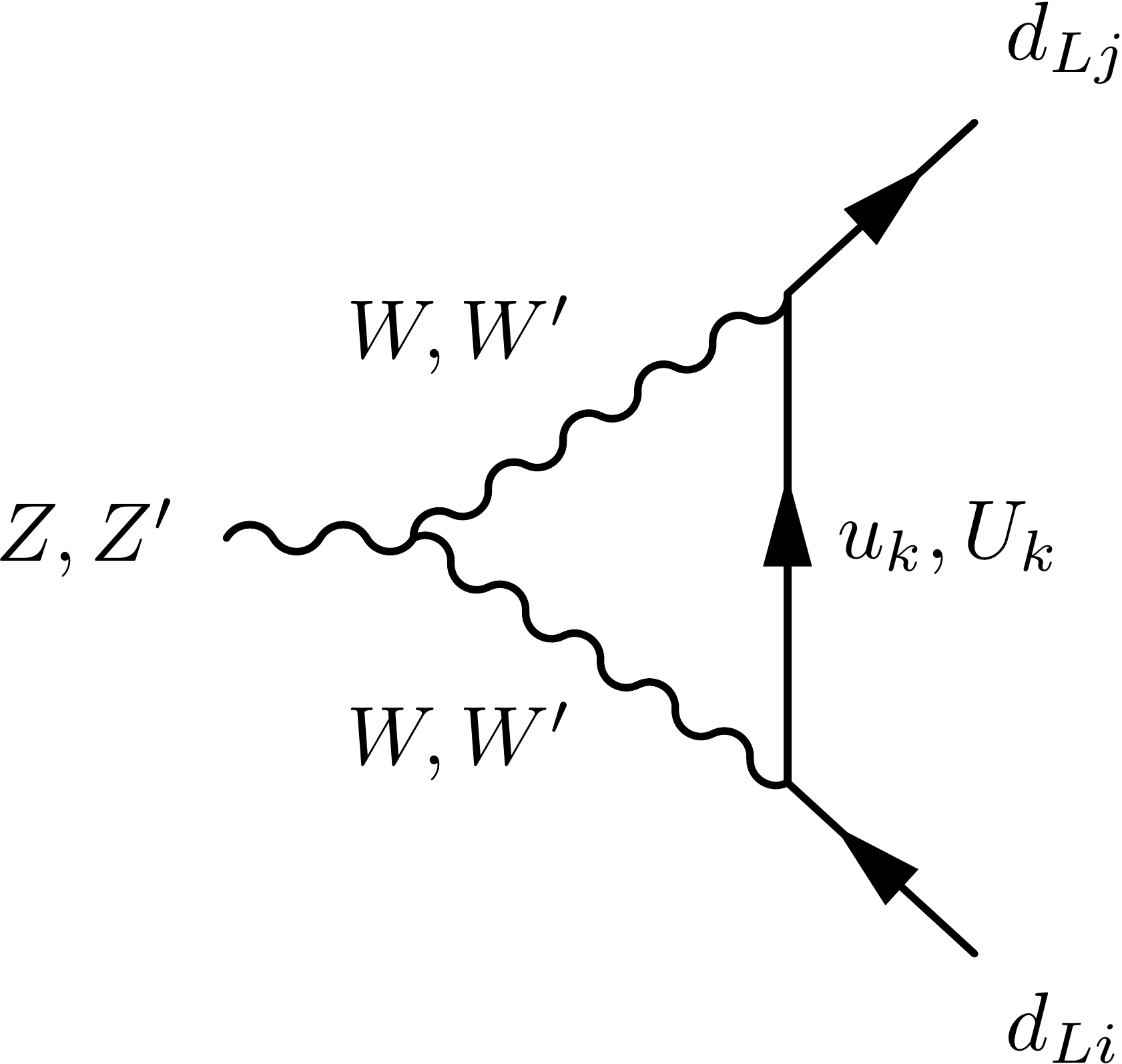}
\hspace{10pt}
\includegraphics[width=5cm,height=5cm]{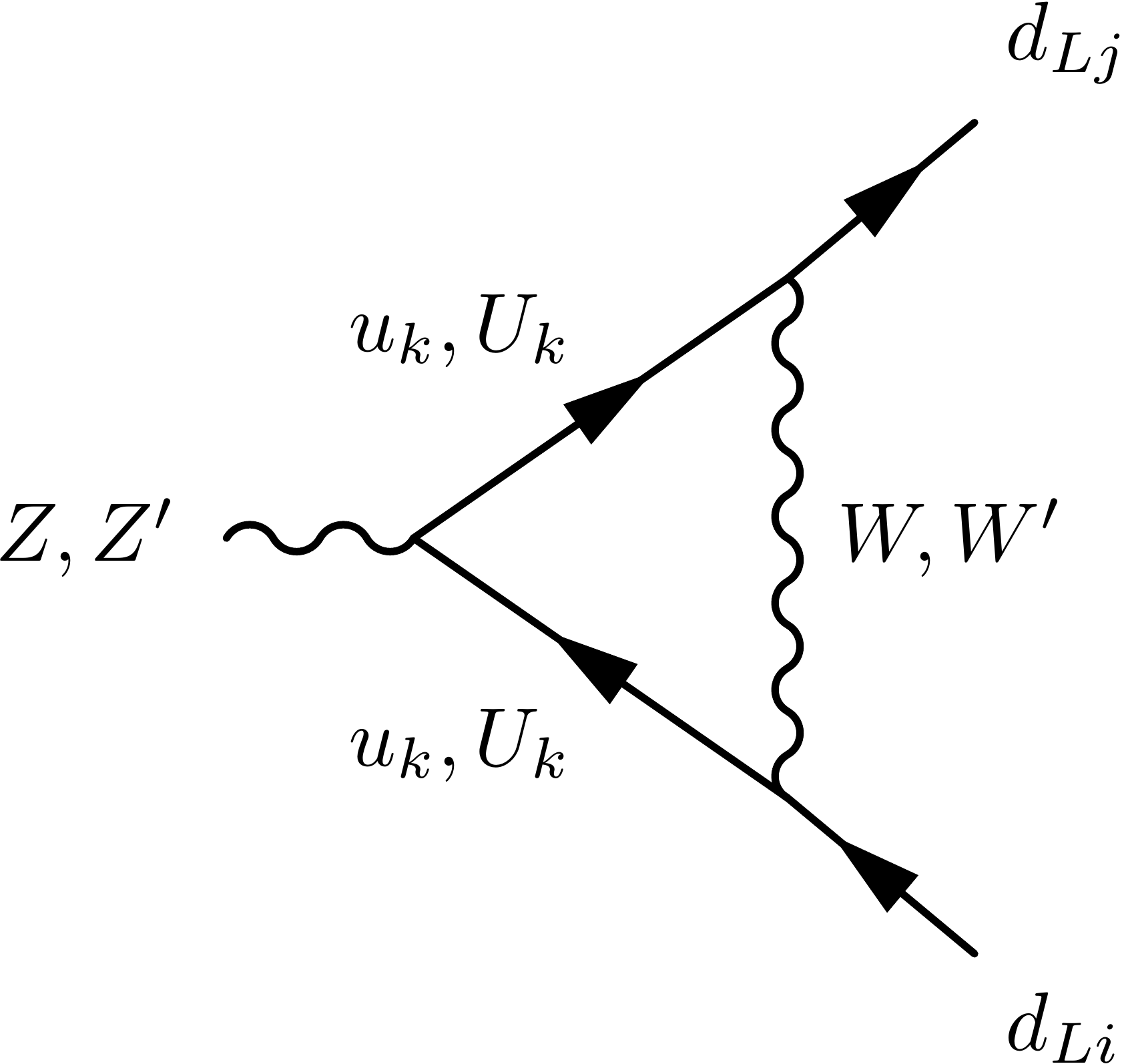}
\caption{Vertex diagrams contributing to the processes $Z \to d_i \bar{d}_j$ and $Z' \to d_i \bar{d}_j$. Each diagram is summed over the
internal $u_k$ and $U_k$ flavors. Due to ideal delocalization, the vertices connecting the heavy $W'$ boson to
light $u\bar{d}$ quark pairs are absent.}
\label{fig:Zvertex}
\end{figure}

\begin{figure}[tb]
\includegraphics[width=5cm,height=5cm]{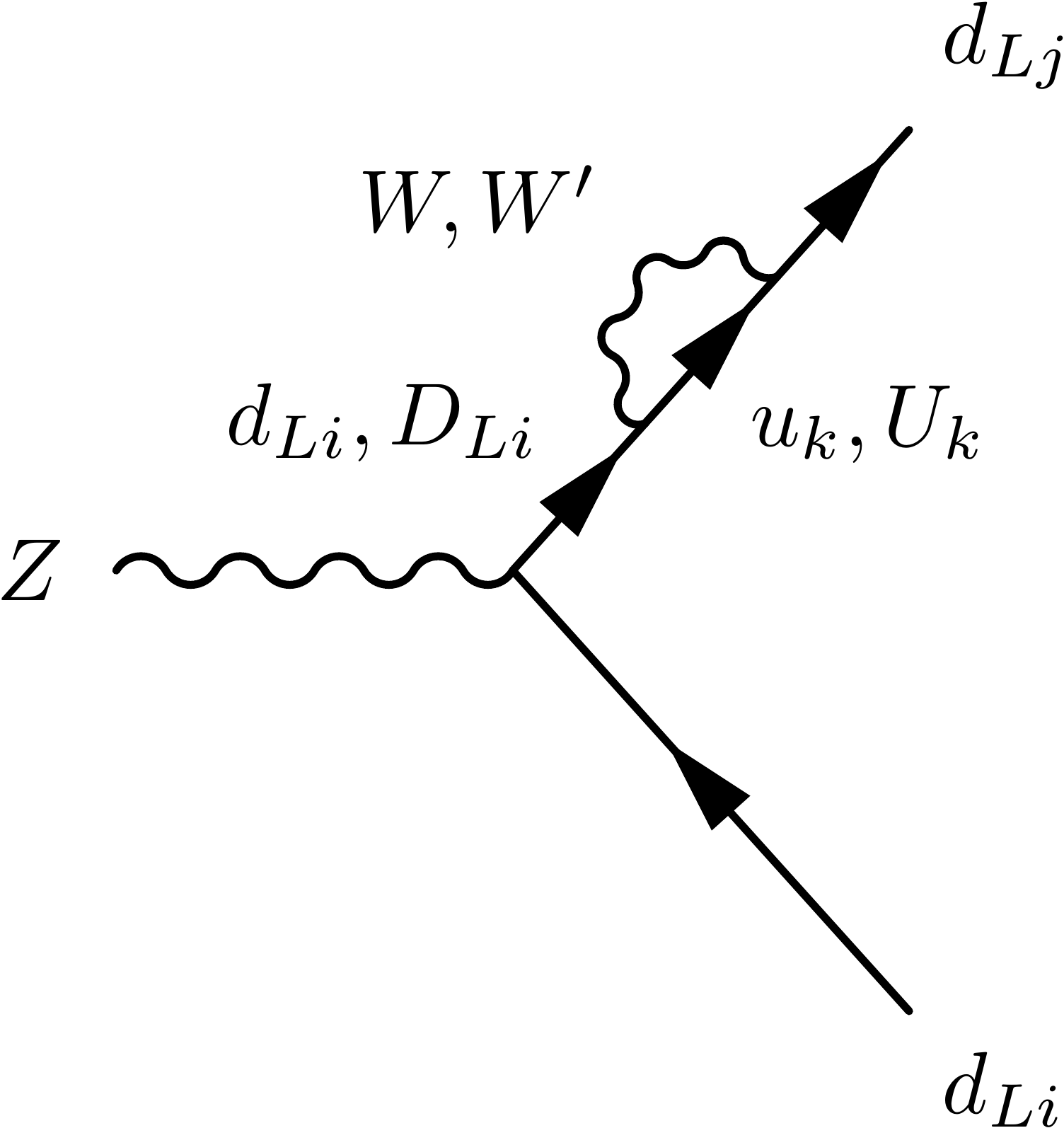}
\hspace{10pt}
\includegraphics[width=5cm,height=5cm]{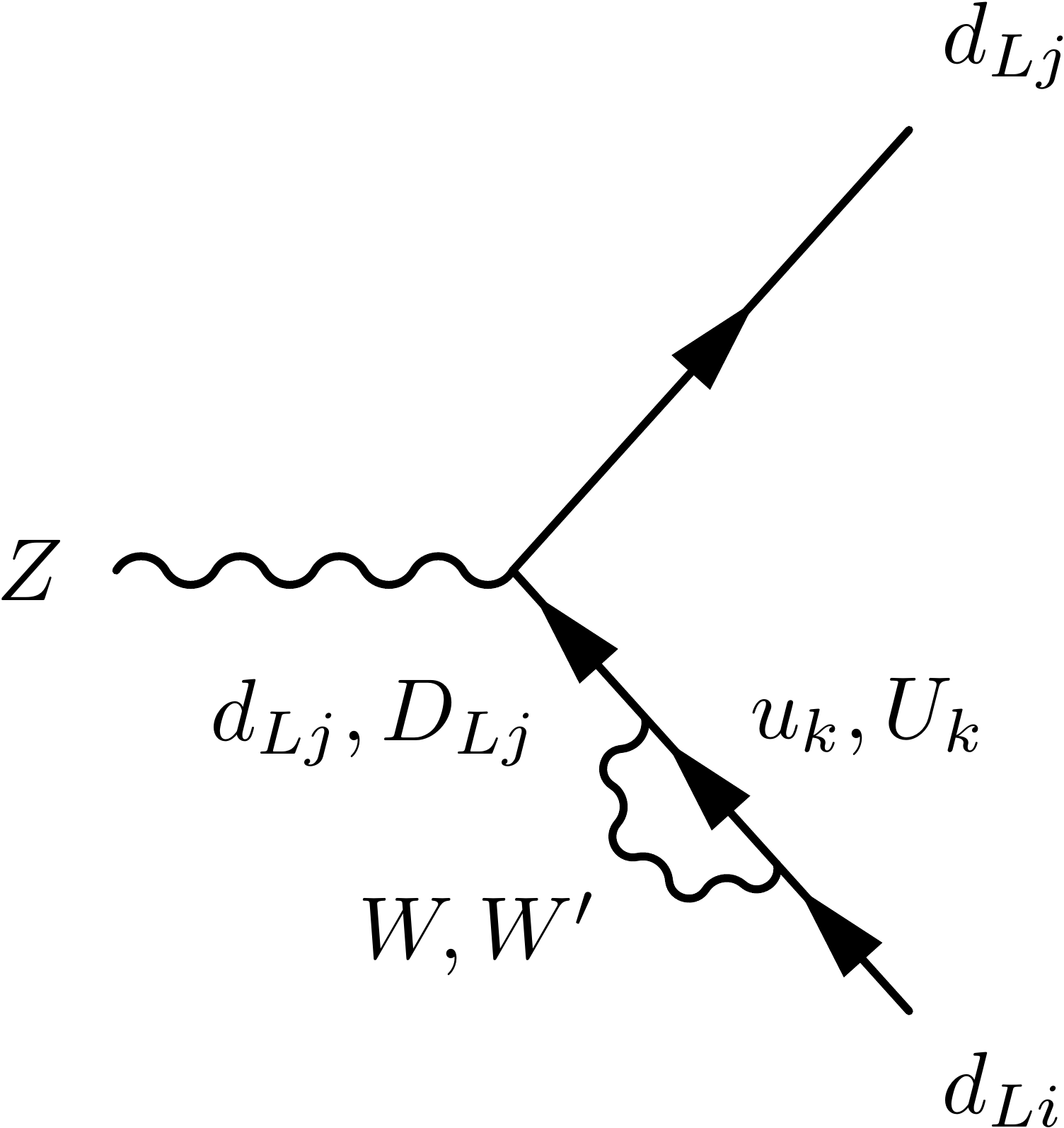}
\caption{Wavefunction renormalization diagrams which must be included in the $Z \to d_i \bar{d}_j$
computation. The analogous $Z'$ contributions are suppressed, relative to the leading vertex contributions.}
\label{fig:Zwavefunction}
\end{figure}

We begin with the calculation of the new contributions to the process $Z \to \bar{f} f'$ in the three-site model. All contributions in the three-site model are shown in Fig. \ref{fig:Zvertex}, though those involving only light particles ({\it i.e.}, those not involving
either the heavy $W'$ or $Z'$ gauge bosons, or the heavy quarks) just reproduce the standard model results. In addition,
one must properly account for the wavefunction corrections illustrated in Fig. \ref{fig:Zwavefunction}. We have performed these
calculations in 't-Hooft-Feynman gauge in the three-site model (the appropriate Feynman rules can be extracted from
references \cite{SekharChivukula:2007ic,Kurachi:2010fa}), but the result is easily understood in terms of the
effective Lagrangian/renormalization group calculation of the flavor non-universal contributions to the $Z \to b{\bar b}$
branching ratio discussed in \cite{Abe:2009ni}. 

\begin{figure}[tb]
\includegraphics[width=7cm,height=3cm]{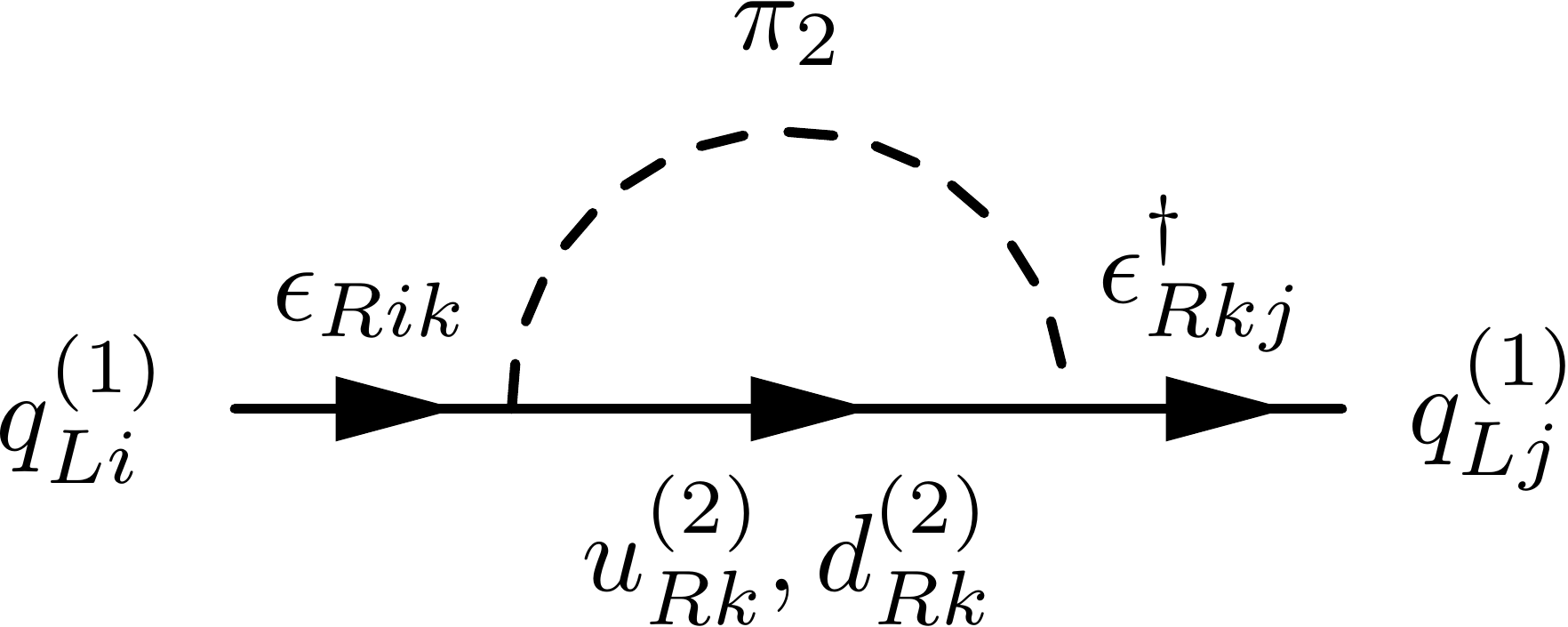}
\caption{Wavefunction renormalization that results in the flavor-dependent running
of the parameter $\epsilon_L$ in the effective theory valid in the energy range
below the cutoff scale and above the masses of the heavy
fermions.  This running yields the renormalization group equation 
(\protect\ref{eq:runningepsilon}). Note that
that the $q^{(1)}_L$ and $u^{(2)}_R$, $d^{(2)}_R$ are the site 1 and 2 gauge-eigenstate fermion
fields of Eq. (\protect\ref{eq:threesite}). In 't-Hooft-Feynman gauge, the leading
contribution comes from  $\pi_2$, the unphysical Nambu-Goldstone Boson in the non-linear sigma model field $\Sigma_2$ of
Eq. (\protect\ref{eq:Sigma}). }
\label{fig:wavefunctionren}
\end{figure}

Applying the results of \cite{Abe:2009ni}, we see that the dominant one-loop effect
in $Z\to \bar{f} f'$ is the flavor-dependent running of the effective Lagrangian parameter $\underaccent{\tilde}{\mathsf M}$ from the cutoff, $\Lambda$ (where ideal delocalization and minimal flavor violation are imposed on the effective Lagrangian parameters) to the scale of the heavy fermion masses.  This effect is due to wavefunction renormalization of the site-1 fermion fields $q^{(1)}_L$, Fig.~\ref{fig:wavefunctionren}. Generalizing
the calculations of \cite{Abe:2009ni}, this wavefunction renormalization results in
the running of the parameter $\epsilon_L \epsilon^\dagger_L$ 
\begin{equation}
\mu \frac{d}{d\mu }\left(\epsilon_L \epsilon^\dagger_L\right)
=-\,\frac{2}{(4\pi)^2 f^2_2}\left[
{\cal M}_u {\cal M}^\dagger_u + {\cal M}_d {\cal M}^\dagger_d \right]~,
\label{eq:runningepsilon}
\end{equation}
where ${\cal M}_{u,d}$ are the mass matrices of the light up- and down-quarks. We see that
the flavor transformation properties (Eq. (\ref{eq:spurion})) of the left- and right-hand sides of
this equation match. Note also that the (dominant) contribution illustrated in Fig. \ref{fig:wavefunctionren} arises
from the unphysical Nambu-Goldstone boson, $\pi_2$, of Eq. (\ref{eq:Sigma}), whose couplings are proportional to the flavor-dependent
parameters $\underaccent{\tilde}{\mathsf m}_{2u,2d}$ and inversely proportional to $f_2$.

Below the
scale of the heavy quark masses, this running ceases. Furthermore, there is a cancellation between 
the vertex and wavefunction
diagrams of Figs. \ref{fig:Zvertex} and \ref{fig:Zwavefunction} because the $SU(2)_0$ global
symmetry to which the $Z$ is largely coupled, is conserved (up to corrections suppressed by electroweak couplings). Denoting the scale of the heavy fermion masses 
by $M$, {\it c.f.} Eq. (\ref{eq:defM}),  we see that the chiral-logarithmic correction to the parameter $\epsilon_L \epsilon^\dagger_L$
is given by
\begin{equation}
\Delta (\epsilon_L \epsilon^\dagger_L)  = \frac{1}{(4\pi)^2 f^2_2}
\left[
{\cal M}_u {\cal M}^\dagger_u + {\cal M}_d {\cal M}^\dagger_d \right]
\log \frac{\Lambda^2}{M^2}~.
\label{eq:deltaepsilonL}
\end{equation}
As usual, from Eqs. (\ref{eq:diagonal-down-masses} -- \ref{eq:diagonal-up-masses}), the first
term gives rise to flavor-changing {\it down} quark couplings while the second to flavor-changing
{\it up} quark couplings. In the case of $s$ and $d$ quarks, for example, from Eq. (\ref{eq:jLcurrent})
we see that the running from the cutoff $\Lambda$ to the scale $M$ of the heavy
quark masses yields the flavor-changing $Z$-boson coupling
\begin{equation}
(g^Z_{\bar{d} s})_{3-site} = \frac{e}{2 (4\pi)^2 \sin\theta_W \cos\theta_W} \frac{f^2_1 f^2_2}{(f^2_1+f^2_2)^2}   \ln (\frac{\Lambda^2}{M^2})   \sum_u
\frac{V^*_{ud} m^2_u V_{us}}{v^2}~,\label{eq:Zsd3site}
\end{equation}
where we have used Eq. (\ref{eq:GF}) to relate the result to $v$. The formulae for the other quarks is similar, with 
the appropriate replacements dictated by the form of $\Delta (\epsilon_L \epsilon^\dagger_L)$ and ${\cal M}_{u,d}$.

By comparison, the corresponding standard model result \cite{Inami:1980fz} is
\begin{equation}
(g^Z_{\bar{d} s})_{SM} = \frac{e}{ (4\pi)^2 \sin\theta_W \cos\theta_W} \sum_u
\frac{V^*_{ud} m^2_u A(m_u,M_W) V_{us}}{v^2}~,\label{eq:ZsdSM}
\end{equation}
where
\begin{align}
A(m_u,M_W) & = \frac{M^2_W(2M^2_W+3m^2_u)}{(m^2_u-M^2_W)^2}\log\left(\frac{m^2_u}{M^2_W}\right)+\frac{m^2_u-6M^2_W}{m^2_u-M^2_W}~,\\
& \to \begin{cases}
\hspace{+20pt} 1 & m_u \gg M_W\\
 2 \log\frac{m^2_u}{M^2_W} + 6 & m_u \ll M_W\,.
\end{cases}
\end{align}
Comparing Eqs. (\ref{eq:Zsd3site}) and (\ref{eq:ZsdSM}), we see that the new three-site model contributions
are, at most, a small fraction of the corresponding (electroweak penguin) standard model result. 
Since the standard model itself yields $Z$-penguin amplitudes too small to be unambiguously observed to date, 
either at the $Z$-pole or in meson decays,  these chiral logarithmic corrections arising from the three-site model are consistent with experiment.

\subsection{$Z' \to \bar{f} f'$}

\begin{figure}[tb]
\includegraphics[width=5cm,height=5cm]{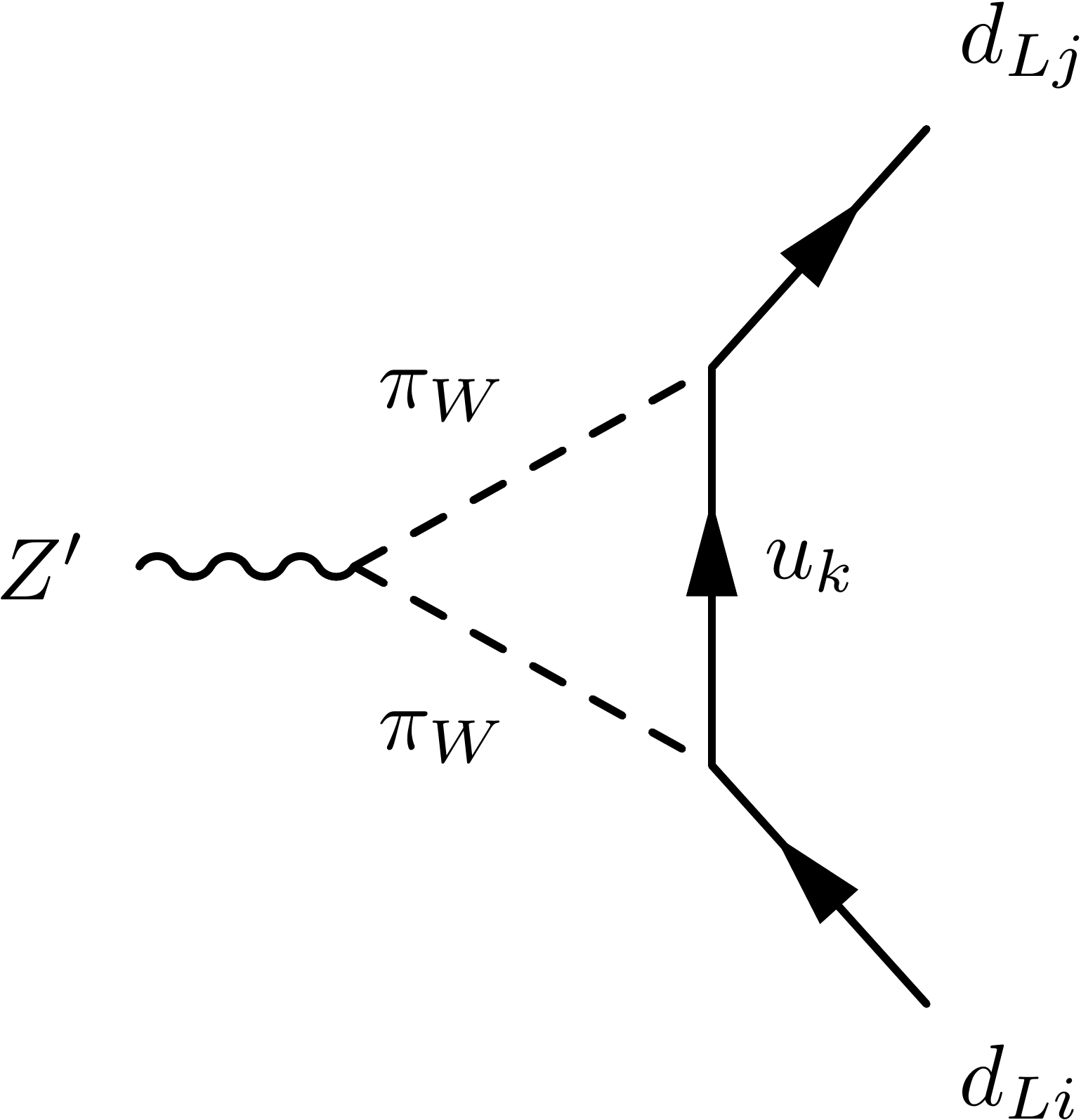}
\caption{$Z'$ flavor-changing vertex renormalization arising in the effective theory valid in the energy range below the
scale of the heavy fermions $M$ and above the weak boson scale $M_W$. In 't-Hooft-Feynman gauge the
leading contribution comes from the $\pi_W$ field,
the unphysical Goldstone boson eaten by the mass-eigenstate $W$-boson. The $\pi_W$
fields couple proportional to the quark masses, as shown
in the effective Lagrangian of Eq. (\protect\ref{eq:Efflag}) and as required by the usual electroweak Ward identities. }
\label{fig:Zprime-vertex}
\end{figure}

Next, for completeness,  we consider flavor changing couplings of the heavy $Z'$ at one-loop.  The form and
size of these couplings illustrate the principles of minimal flavor violation and effective field theory
we have discussed in the previous section. However, in practice, these
couplings are of little phenomenological consequence: because of ideal delocalization, Eq. (\ref{eq:ideal-delocalization}),
the only couplings to light fermions are the small hypercharge-related terms in Eq. (\ref{eq:newheavycouplings}).
Therefore, these couplings cannot appreciably contribute to processes such as $B_{s,d} \to \mu^+ \mu^-$.

Calculating the diagrams shown in Figs. \ref{fig:Zvertex} and \ref{fig:Zwavefunction}, we find the leading
flavor-changing contributions
\begin{equation}
g^{Z'}_{\bar{s}d} = -\,\frac{\tilde{g}}{2(4\pi)^2} \left(\sum_u \frac{V^*_{us} m^2_u V_{ud}}{v^2} \right)
\left[\frac{v^2}{f^2_2} \log\frac{\Lambda^2}{M^2} + \log \frac{M^2}{m^2_u}\right]~,
\label{eq:gzprimesd}
\end{equation}
where, for illustration, we have considered the $\bar{s}d$ coupling; the generalization to other
quark flavors is dictated by the minimal flavor-violating structure. The origin of the two terms in Eq. (\ref{eq:gzprimesd})
is rather different. The first term (proportional to $\log(\Lambda^2/M^2)$) exhibits how the running
of $\epsilon_L$ in Eq. (\ref{eq:deltaepsilonL}) affects the $Z'$ couplings shown in Eq. (\ref{eq:newleft}).
The second term, as indicated by the presence of  $\log(M^2/m^2_u)$, arises in the effective theory between the scale of
the heavy fermions ($M$) and the quark mass (here we assume $m_u =m_t \gg M_W$) in the loop shown in Fig.
\ref{fig:Zprime-vertex}. 

In the end, we conclude that there are no phenomenologically significant flavor-changing effects in $Z'$ couplings at this order.
As noted above, ideal delocalization eliminates any tree-level flavor-diagonal $Z'$ coupling to light  fermions, 
While the presence of the large coupling $\tilde{g}$ in the one-loop result of Eq. (\ref{eq:gzprimesd}) is tantalizing, that enhancement
is cancelled in any low-energy process by the suppression from inverse powers of the $Z'$ mass.  Hence, there are no appreciable 
$Z'$-exchange contribution to $\Delta F=1$ processes.  In principle, $Z'$-exchange
contributions to $\Delta F=2$ processes are possible -- but these are two-loop effects which are
substantially smaller than the one-loop standard model ``box-diagram" contributions, as we will discuss in Section \ref{sec:delf2}.

\subsection{$b \to s \gamma$}

In the subsections above, we have focused on flavor-changing couplings of the $Z$ and $Z'$ bosons. Notably, we saw that
the minimal flavor violation of the three-site model implies that the leading new-physics effects are 
confined to the left-handed sector, just as in the standard model. In contrast, gauge invariance
and minimal coupling insure that the chirality preserving couplings of the photon are flavor-diagonal. Instead,
the leading operator for the phenomenologically relevant radiative decay $b\to s \gamma$ has the form \cite{Grinstein:1987vj}
\begin{equation}
{\cal H}_{eff} = -\,\frac{4 G_F\, A(m_t,M_W)}{\sqrt{2}} V^*_{ts} V_{tb}
\left( \frac{e}{16\pi^2} m_b(\bar{s}_L \sigma^{\mu\nu} b_R)F_{\mu\nu}\right)~,
\end{equation}
where, to leading order in the standard model \cite{Inami:1980fz}
\begin{equation}
A(m_t,M_W) = \frac{3x^3-2x^2}{4(x-1)^4}\log x + \frac{-x^3+5x^2+2x}{8(x-1)^3};
\hspace{1cm} x=\frac{m^2_t}{M^2_W}~.
\end{equation}

New contributions to this process arise in the three-site model from the presence of  {\it right-handed}
couplings of the W to $b$-quarks (see  Eq. (\ref{eq:newright}) and \cite{SekharChivukula:2006cg}), as well as from the presence of new heavy particles in the loop.
These contributions have been studied in detail in \cite{Kurachi:2010fa}, for the special case $f_1=f_2=\sqrt{2}v$. Their results show that the new contributions
are only of order 10\% of the standard model contribution for the preferred range of $\epsilon_{Rt} < 0.3$ \cite{SekharChivukula:2006cg},  and that including the contributions from the three-site model tends to {\it improve} the consistency with the 
experimental results.  Varying away from the point $f_1 = f_2$ will increase the masses of the additional particles,
decreasing the size of the new three-site corrections. At the very least, the three-site model's prediction for the rate of $b\to s \gamma$ will be as consistent with experimental
data as that of the standard model.

\section{$\Delta F=2$ Processes at One Loop}
\label{sec:delf2}

 \begin{figure}[b]
 \begin{center}
 \includegraphics[angle=0, scale=0.75]{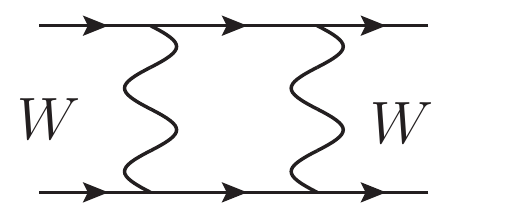}
 \includegraphics[angle=0, scale=0.75]{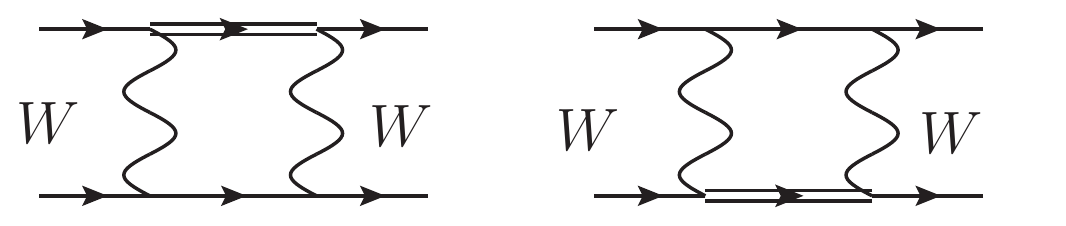}
 \includegraphics[angle=0, scale=0.75]{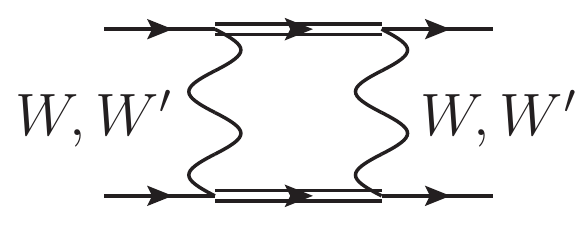}
 \caption{Diagrams that give dominant contributions to $\Delta F =2$
  processes. The single lines and the double lines represent the standard model
  and KK fermions, respectively. Because of ideal delocalization, the $W'$ boson
  does not couple to two light fermions -- and therefore only contributes in diagrams involving
  two heavy intermediate states. The first diagram, including only light standard model states,
  receives non-standard contributions in the three-site model only to the extent that the weak gauge couplings
  differ from their standard model equivalents at ${\cal O}(x^4)$.}
\label{fig:diagrams}
 \end{center}
 \end{figure} 

 \begin{figure}[t]
 \begin{center}
 \includegraphics[angle=0, scale=0.2]{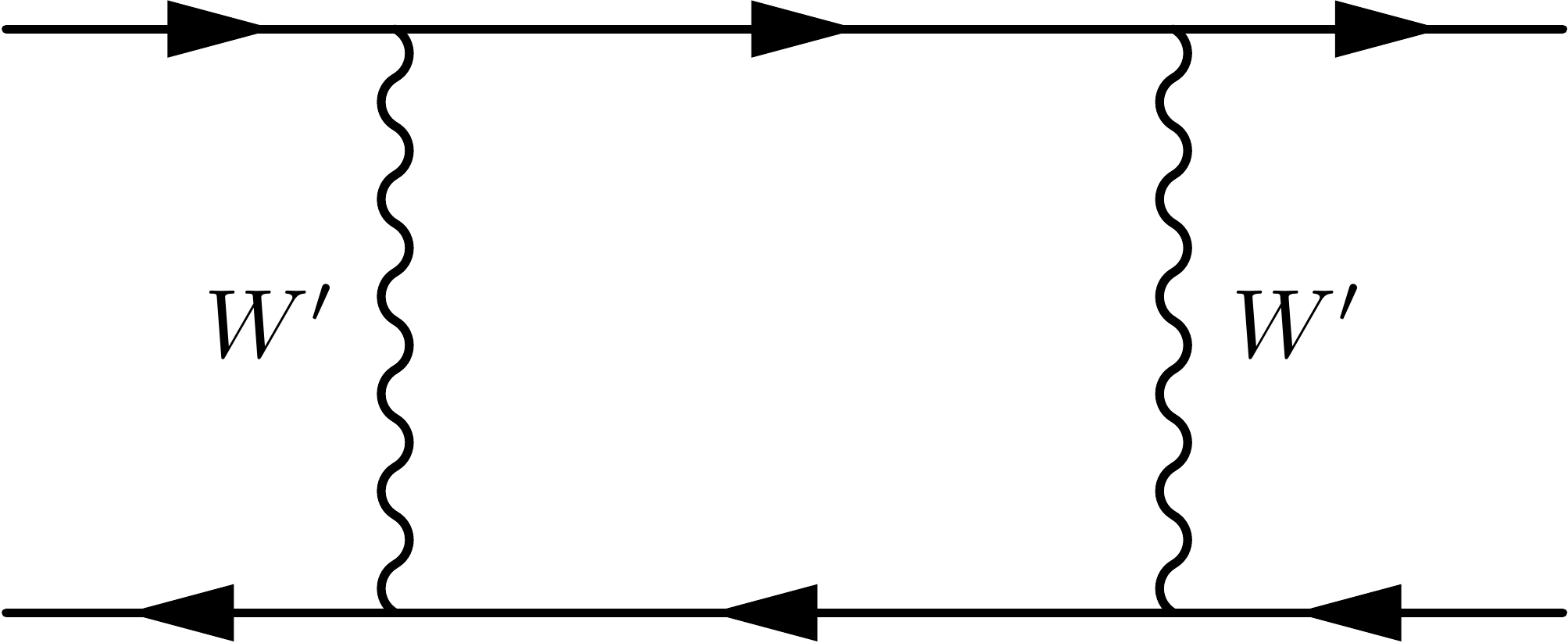}
 \hskip25pt
 \includegraphics[angle=0, scale=0.2]{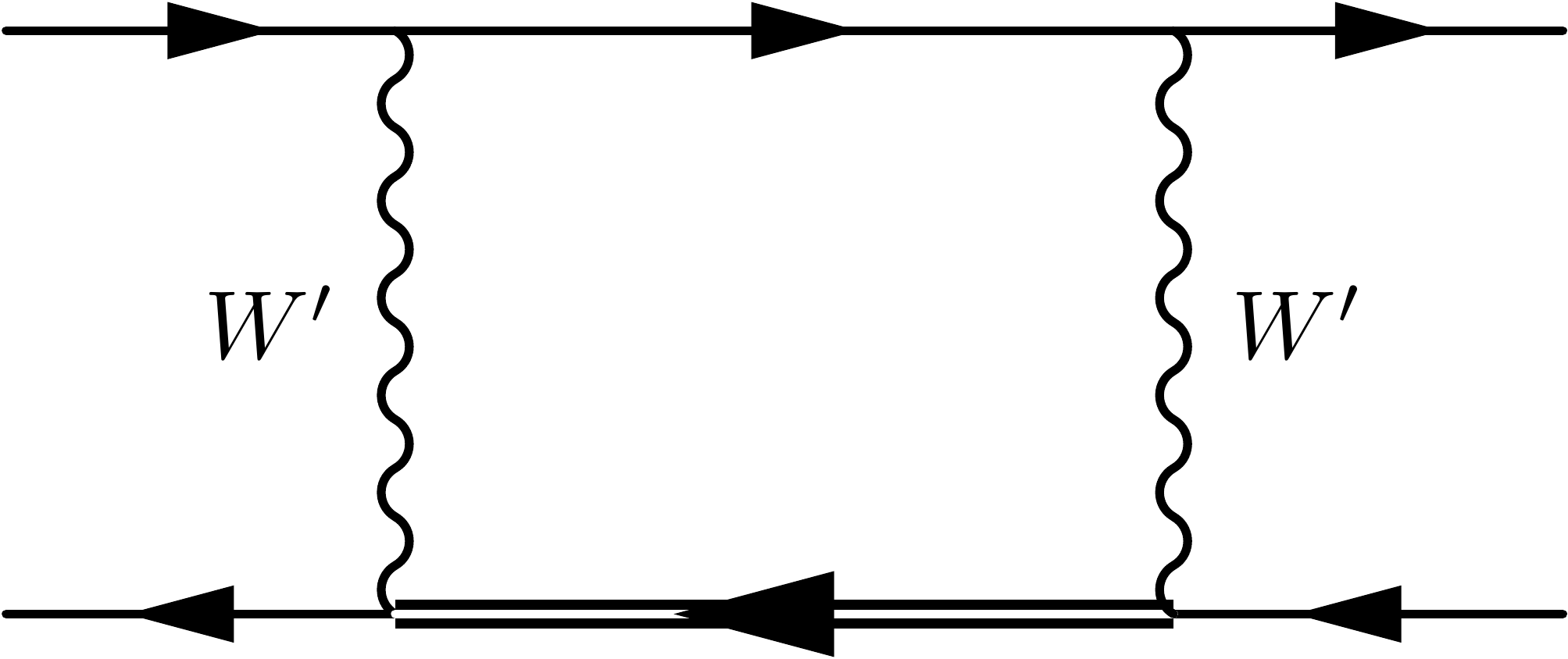}
\hskip25pt
 \includegraphics[angle=0, scale=0.2]{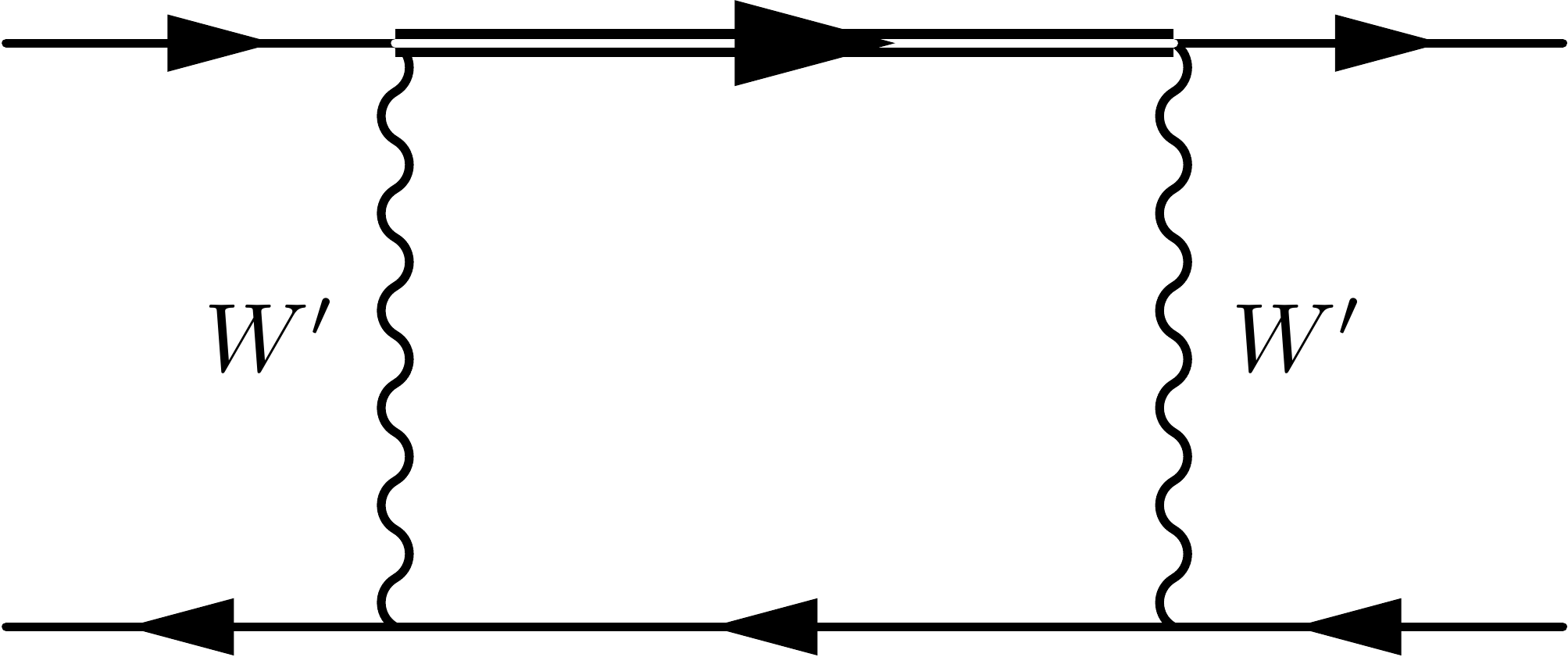}\\ 
  \includegraphics[angle=0, scale=0.2]{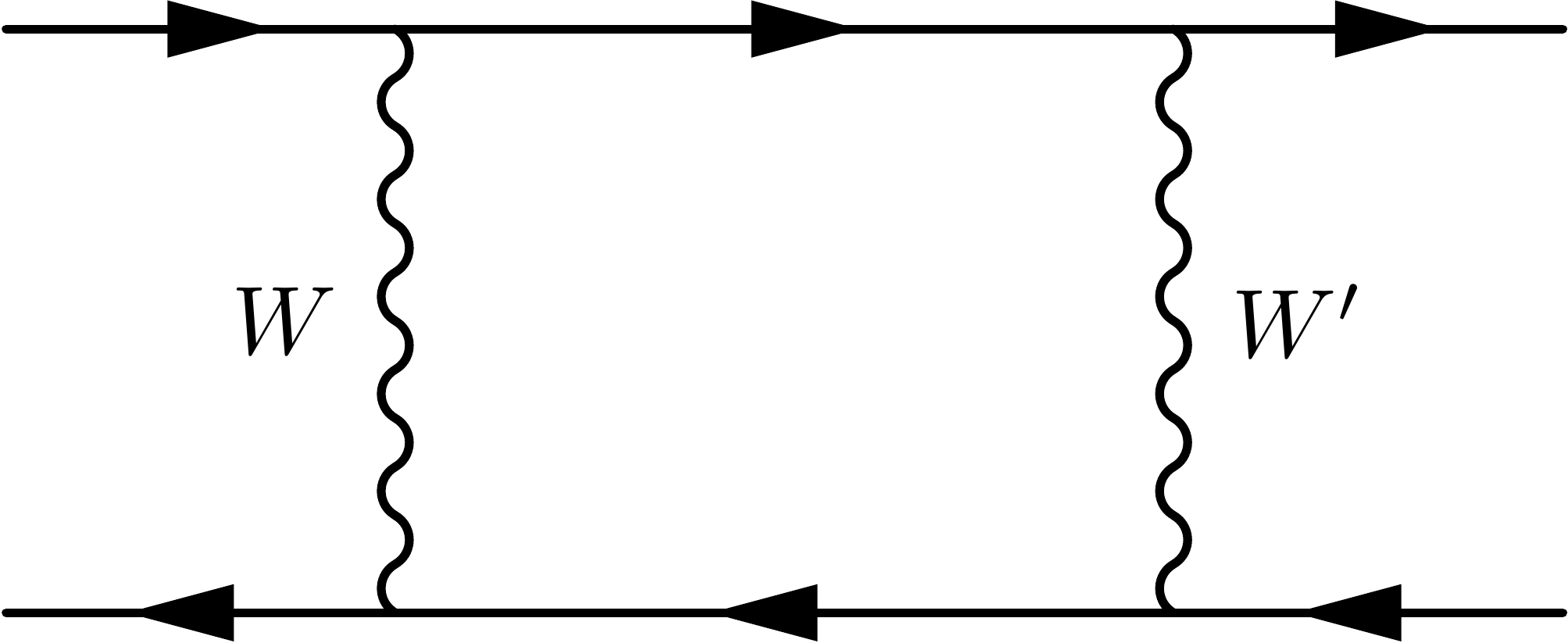}
 \hskip25pt
 \includegraphics[angle=0, scale=0.2]{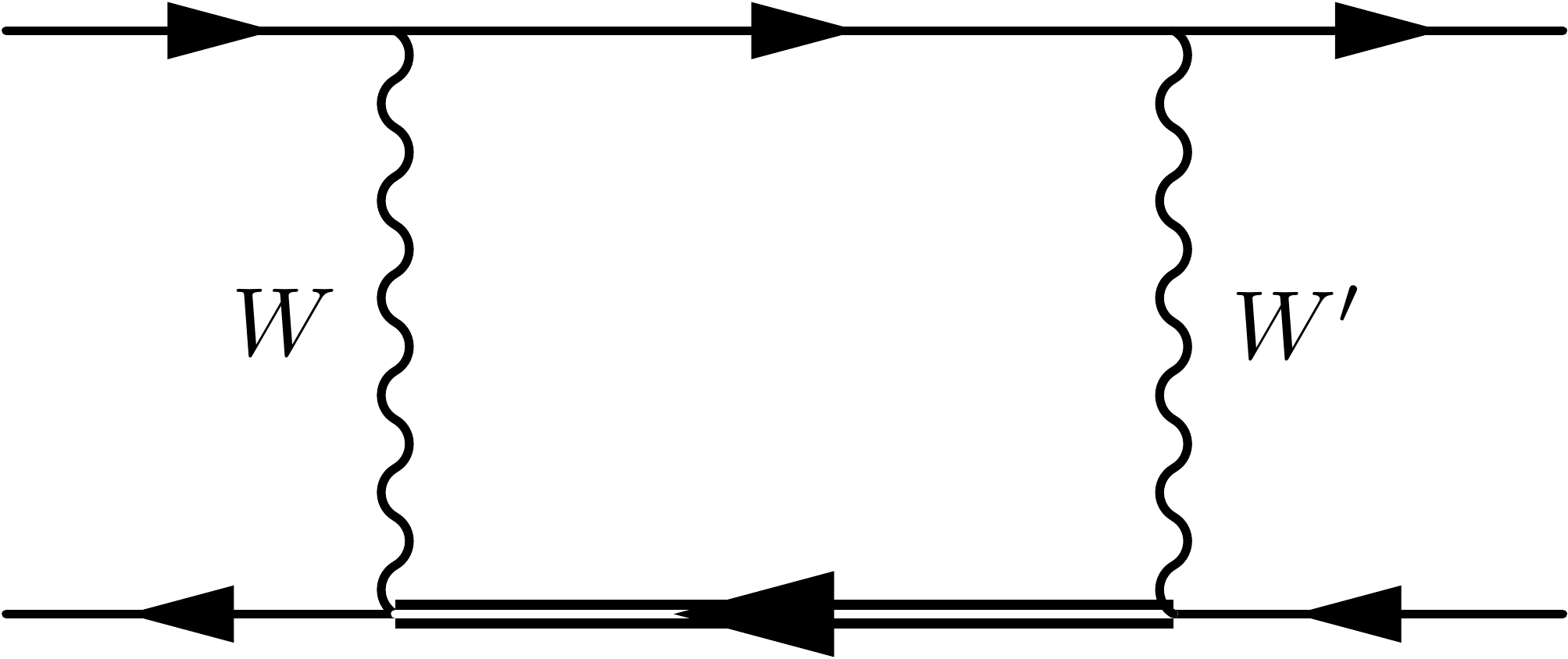}
\hskip25pt
 \includegraphics[angle=0, scale=0.2]{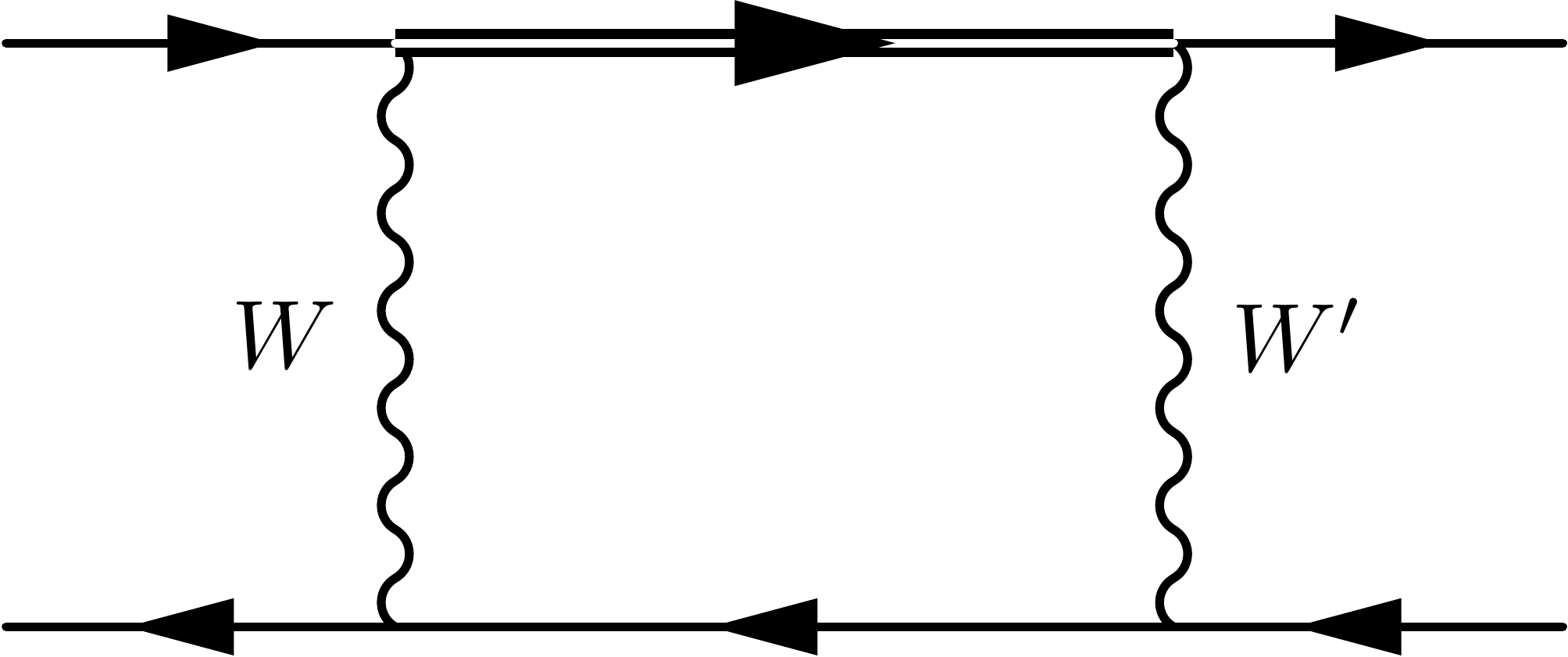}\\ 
  \includegraphics[angle=0, scale=0.2]{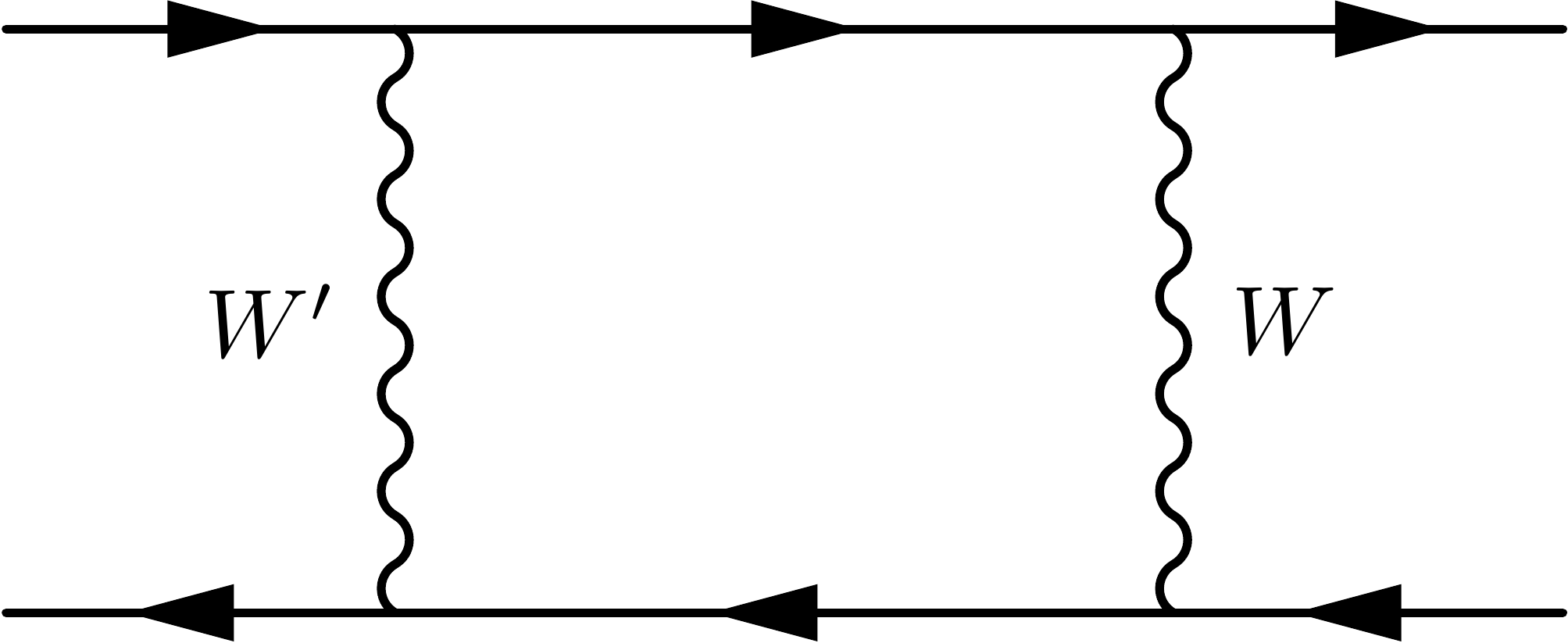}
 \hskip25pt
 \includegraphics[angle=0, scale=0.2]{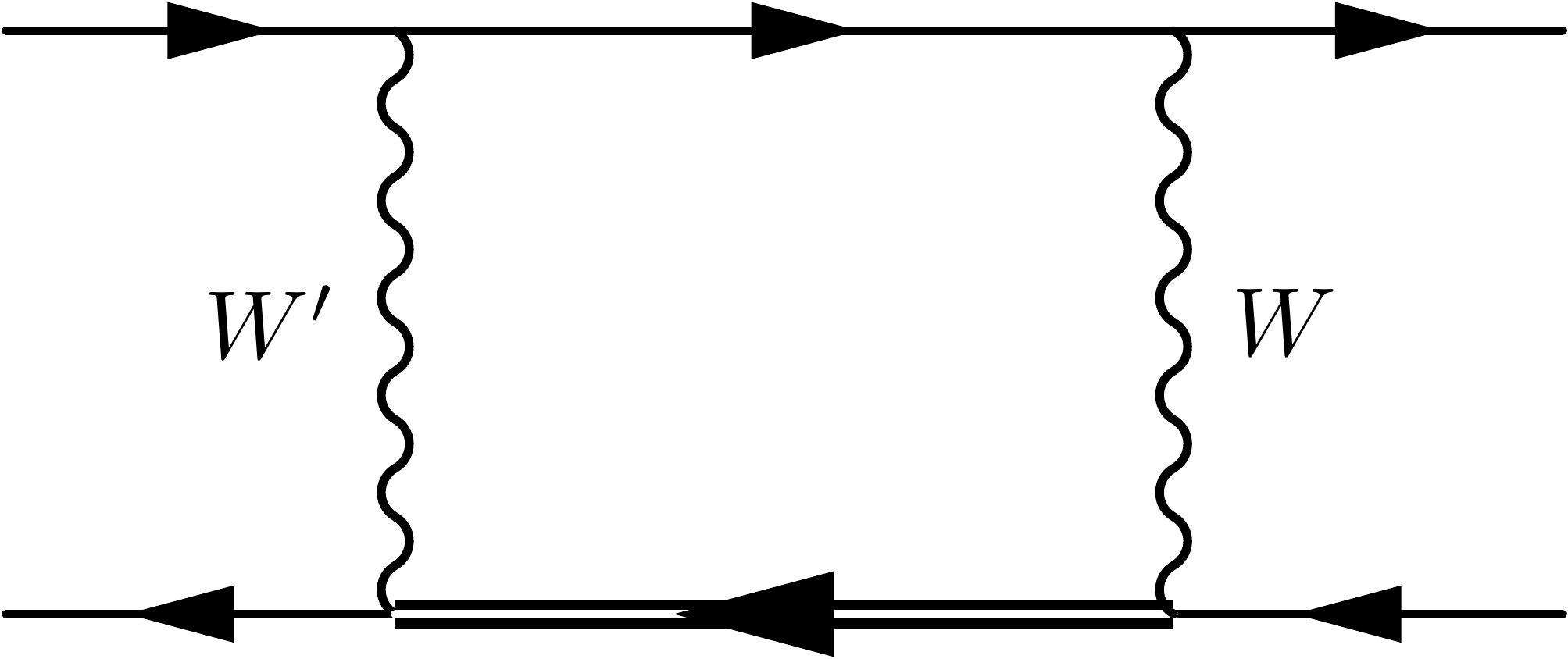}
\hskip25pt
 \includegraphics[angle=0, scale=0.2]{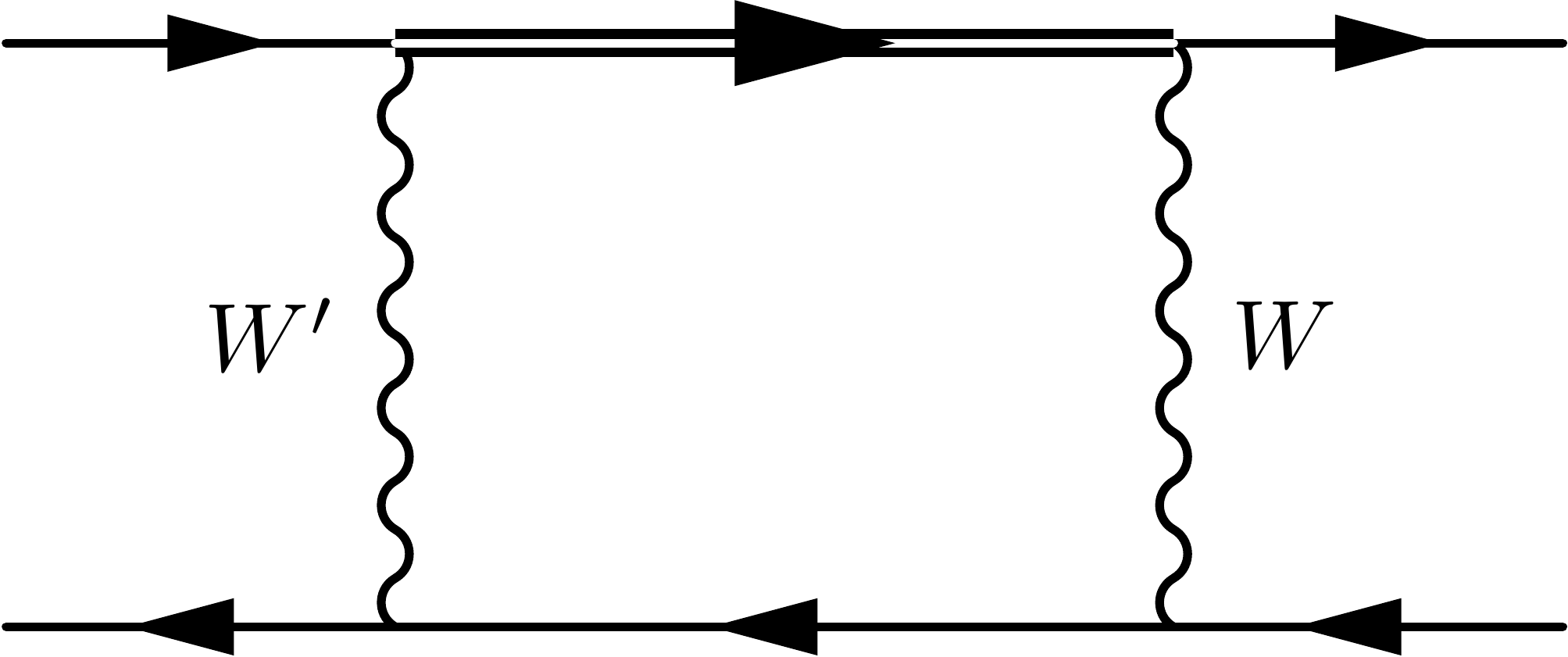}
 \caption{Diagrams that are absent from the calculation of $\Delta F = 2$ processes in the three-site model, due to ideal delocalization. The single and double lines represent
 standard model and KK fermions respectively.  As described in the text, the combination of ideal delocalization and minimal
 flavor violation implies that all $\Delta F=2$ effects are suppressed by $(m/M)^2$, where $m$ and $M$ are
 the masses of the standard model and KK fermions respectively. }
\label{fig:wpdiagrams}
 \end{center}
 \end{figure} 

In this section, we extend the analysis of the previous section to study the chiral-logarithmically enhanced corrections to $\Delta F = 2$ processes in the three-site model.
We  show that the combination of minimal flavor violation and ideal fermion delocalization ensures that both the one-loop corrections from $\Delta F = 2$ box diagrams and the 
two-loop corrections from $\Delta F = 1$ vertices are small compared to similar corrections in the standard model.

The contributions to $\Delta F=2$ processes in the three-site model are shown in Fig. \ref{fig:diagrams}.
The contribution from the first diagram corresponds to those in the standard model. 
Since the couplings of the $W$ in the three-site model agree with their standard model counterparts up to corrections ${\cal O}(x^4) \laem 10^{-3}$,
this diagram essentially reproduces the standard model contribution.  In particular, GIM cancellations imply that all contributions involving light fermions
are suppressed by {\it four} powers of the light up-quark masses. Furthermore, because of ideal delocalization, the diagrams shown in Fig. \ref{fig:wpdiagrams} are absent.
The absence of the first (upper left-most) diagram insures that there are no new ``long-distance" contributions in the
three-site model, nor other new contributions depending on light quark masses but not heavy KK quark masses. 

Returning to Fig. \ref{fig:diagrams}, we recall that $\underaccent{\tilde}{\mathsf M}$ and $\underaccent{\tilde}{\mathsf m}_1$ are flavor-diagonal
at tree-level, and that the masses of the heavy KK fermions are approximately degenerate -- with deviations
 proportional to the corresponding light fermion masses \cite{SekharChivukula:2006cg}. Hence
GIM cancellation in the {\it heavy} fermion sector implies that  contributions from the last three diagrams in
Fig. \ref{fig:diagrams} are suppressed by $m^2_q/M^2\laem {\cal O}(10^{-3})$, where $m_q$ is a mass of a light quark and
$M$ is the mass of the KK fermions. To summarize, the combination of ideal delocalization and minimal
flavor violation insure that the {\it new} contributions to $\Delta F=2$ box diagrams in the three-site model
are smaller than or of order {\it two-loop} corrections to the same $\Delta F = 2$ processes in the standard model -- and hence
are not phenomenologically excluded. 

 \begin{figure}[]
 \begin{center}
 \includegraphics[angle=0, scale=0.2]{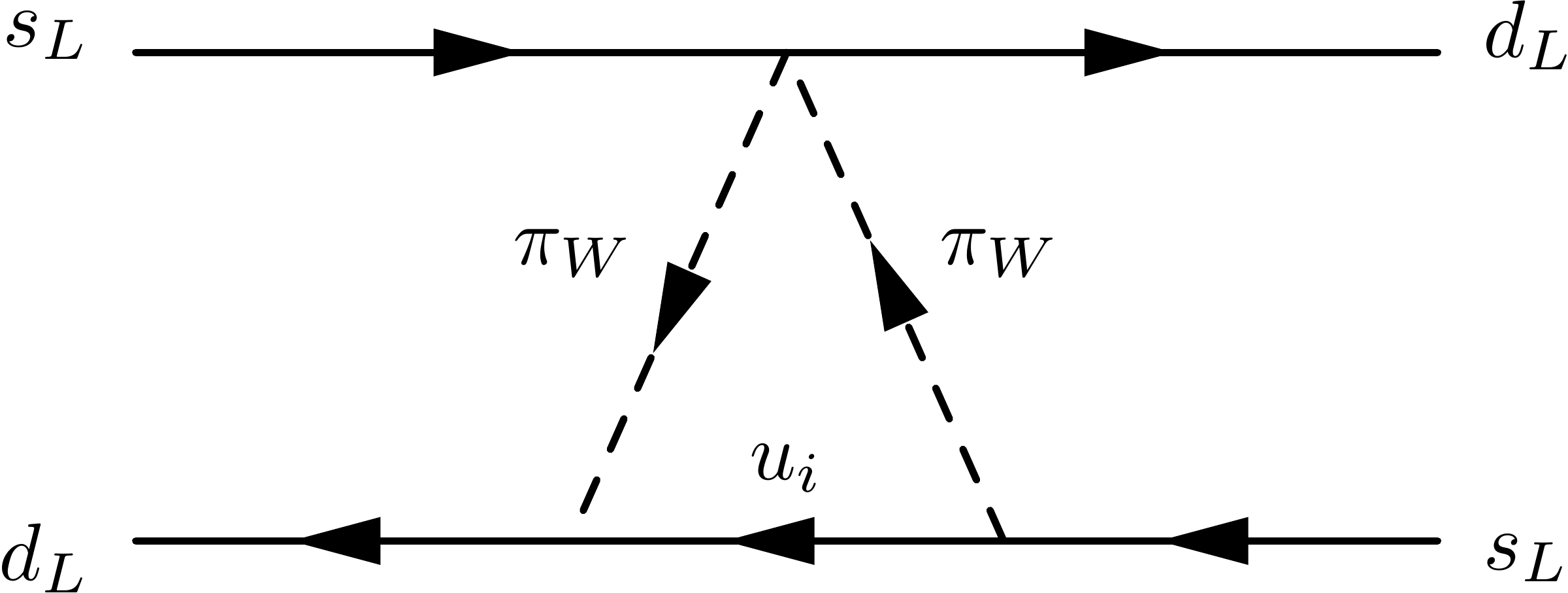}
 \hskip25pt
 \includegraphics[angle=0, scale=0.2]{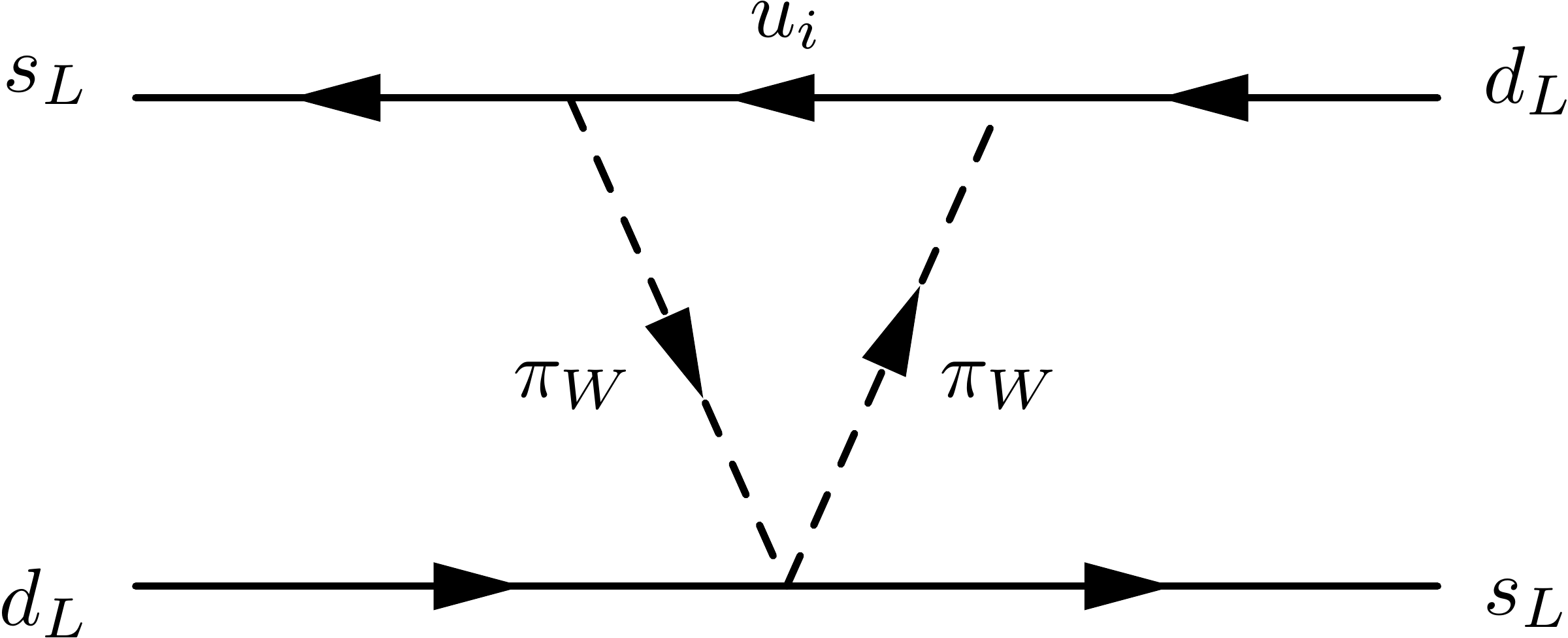}
 \caption{Diagrams that yield the chiral logarithmically-enhanced
 three-site model corrections to $\Delta S=2$ processes in the effective theory, Eq. (\protect\ref{eq:Efflag}), including the flavor non-universal
 corrections to $\epsilon_L$ shown in Eq. (\ref{eq:nonuniversal}), . Here $\pi_W$ are the 't-Hooft-Feynman gauge 
 unphysical Goldstone Bosons eaten by the $W$ boson. As discussed in the text, though enhanced by
 $\log(M^2/M^2_W)$, all new three-site contributions are even more deeply suppressed by $m^2_q/M^2$.}
\label{fig:effective-box}
 \end{center}
 \end{figure} 

These points can be illustrated in more detail by considering the leading, chiral-logarithmically enhanced, three-site box diagram
contributions, which arise from the second and third diagrams in Fig. \ref{fig:diagrams}. These contributions can be
described in effective field theory as follows. The rotations defining the left-handed fermion mass eigenstate fields, 
Eq. (\ref{eq:light}), are, to leading order, proportional to $\epsilon_L$ and therefore flavor-diagonal. The largest
flavor non-diagonal contributions, which can be obtained by diagonalizing ${\cal M}^\dagger {\cal M}$ to higher order (with
${\cal M}$ defined in Eq. (\ref{eq:quarkmasses})), correspond to modifying\footnote{There is
an analogous shift to $\varepsilon_R$ proportional to $\epsilon_L^\dagger \epsilon_L$ --- however, since $\epsilon_L$ is
flavor-diagonal at tree-level, these corrections are flavor-universal.}  $\epsilon_L$ 
\begin{equation}
\Sigma_1 \epsilon_L \to \Sigma_1 \epsilon_L \cdot \left( {\cal I} - {\mathsf M} \Sigma_2 \varepsilon^\dagger_R \varepsilon_R \Sigma^\dagger_2 {\mathsf M}^{-1}\right)~.
\label{eq:nonuniversal}
\end{equation}
Note that these corrections are consistent with the spurion transformations of Eq. (\ref{eq:spurion}). Plugging
this correction into the left-handed couplings on the second line of Eq. (\ref{eq:Efflag}) yields, in 't-Hooft-Feynman
gauge, flavor-changing couplings between the left-handed mass-eigenstate quarks and the Goldstone bosons
eaten by the $W$ boson. These flavor-changing couplings are proportional to $(\epsilon_L \epsilon_{qR})^2
= (m_q/M)^2$, and the overall result (summing over all intermediate heavy-quark flavors) must include the
appropriate CKM mixing matrix elements.

The chiral-logarithmically enhanced three-site box contributions correspond, in the effective theory with heavy KK quarks integrated out, to
the diagrams illustrated
in Fig. \ref{fig:effective-box} (here shown for $\Delta S=2$ processes). Viewing these diagrams as a contribution
to the effective operator
\begin{equation}
{\cal H}^{\Delta S=2}_{eff} = C^1_K (\bar{s}_L \gamma^\mu d_L)(\bar{s}_L \gamma_\mu d_L)~,
\end{equation}
we obtain the leading three-site contribution
\begin{equation}
\left(C^1_K\right)_{three-site} = -\, \frac{\sqrt{2}G_F}{(4\pi)^2}\cdot
\frac{v^2}{M^2} \cdot \left(
\sum_u
\frac{V^*_{ud} m^2_u V_{us}}{v^2}
\right)^2\cdot
\log\frac{M^2}{m^2_W}~.
\end{equation}
Note that this expression exhibits the specific features described above: (a) suppression by four powers of light fermion masses, (b)
as consistent with flavor symmetry, light fermion masses appearing in combination with the usual products of CKM angles,
(c) suppression by the heavy KK fermion masses $M$, and (d) logarithmic enhancement corresponding
to ``running" from $M$ to the $W$ mass. The generalization\footnote{There are also other terms  that are parametrically smaller (e.g. of order $x^4$) but numerically similar in size to those discussed here; since they are also small compared to the standard model contributions, including them would not alter our conclusions.} to other $\Delta F=2$ processes is
straightforward, as determined by flavor symmetry.

Finally, as shown in the previous section, there are no anomalously large $\Delta F=1$
corrections at the one-loop level in the three-site model -- hence, combinations of these $\Delta F = 1$ contributions
produce very small $\Delta F=2$ amplitudes. We conclude that, due to ideal delocalization and
minimal flavor violation, there are no large corrections to $\Delta F=2$ processes in the three-site model.

\section{Conclusions}

In this paper we have explored the flavor structure of the three-site model, and the size of new three-site model contributions to chirality preserving
flavor-changing neutral current processes.
We established the conditions under which the three-site model exhibits minimal or non-minimal flavor violation, and showed 
that experimental bounds on flavor-changing effects constrain the tree-level Lagrangian of the three-site model to a form exhibiting only minimal flavor violation.

Assuming minimal flavor violation at the scale of the ``cutoff", {\it i.e.} the scale of the physics underlying the effective three-site model, we
have computed the chiral logarithmic corrections to chirality-preserving flavor-changing neutral current processes. We have shown that the combination of ideal
delocalization and minimal flavor violation imply that all flavor-changing $\Delta F=1$ neutral current
processes are parametrically the same size as in the standard model, but numerically smaller. In the case of $\Delta F=2$
neutral current processes, the combination of ideal delocalization and minimal flavor violation imply that the three-site model
contributions are smaller than or of order the two-loop corrections to these processes in the standard model. We conclude, therefore, that the three-site model is phenomenologically consistent with experimental data on 
(chirality preserving) flavor-changing neutral current processes.

\bigskip

\begin{acknowledgments}
 RSC and EHS were supported, in part, by the US National Science Foundation under grant PHY-0854889 and also acknowledge the hospitality of the Institute for Advanced Study and the Aspen Center for Physics, where part of this work was completed. M.T.'s work is supported in part by the JSPS Grant-in-Aid for Scientific Research 
    No.20540263 and No.23540298, and he acknowledges the hospitality of the Aspen Center for Physics where part of this work was completed. The figures
    for this paper were drawn using JaxoDraw \cite{Binosi:2003yf} and Feynmf \cite{Ohl:1995kr}.

\end{acknowledgments}


\appendix
\section{Experimental constraints on the form of $\epsilon_L$}

In this Appendix, we establish the experimental constraints on the flavor structure of the charged lepton and quark sectors of the three-site model.  We start by calculating the fermion couplings to the weak gauge bosons in a general framework that does not assume ideal fermion delocalization or lepton universality. We then determine the bounds placed on the  flavor structure by precision electroweak data and studies of flavor-violating processes.  The results demonstrate that the matrices in the tree-level three-site model Lagrangian that govern the delocalization of left-handed quarks ($\epsilon_L$) and left-handed leptons ($\epsilon_{L\ell}$) must be flavor-universal and consistent with ideal fermion delocalization.

\subsection{Electroweak Couplings in the Three-Site Model}
\label{subsec:ewsm}

In order to compare the electroweak couplings of the fermions in the three-site
model with precision electroweak and flavor data, we must first compute the couplings of the fermions to
the $W$- and $Z$-bosons. Unlike the analysis in \cite{SekharChivukula:2006cg}, here we will
{\it not} assume ideal delocalization: instead, we will compute the
couplings for arbitrary values of the delocalization parameter for the left-handed fermions, $\varepsilon^2_L$ and the ratio 
\begin{equation}
x^2=\frac{g^2}{\tilde{g}^2} \approx \left(\frac{f^2_1+f^2_2}{v^2}\right)\left(\frac{M_W}{M_{W'}}\right)^2~, \label{eq:xxdef}
\end{equation}
In other words, rather than imposing the relation in Eq. (\ref{eq:ideal-delocalization}),  we study the degree to which $\epsilon_L \epsilon^\dagger_L$ can deviate from that ideal delocalization value $(\varepsilon^{ideal}_L)^2 \cdot {\cal I}$.

To investigate the electroweak phenomenology of our model,
we display our results in terms of  the charge of the electron (\ref{eqn:edef}) and the ``on-shell" definition of the weak
mixing angle \cite{:2005ema}:
\begin{equation}
\cos^2\theta_W = \frac{M^2_W}{M^2_Z}~.
\label{eq:onshellmixing}
\end{equation}
Diagonalizing the vector-boson mass matrices, applying
the fermion wavefunctions in Eq. (\ref{eq:light}), and rewriting
the results in terms of $e$
 and $\sin\theta_W$, we find\footnote{These expressions are consistent, to the appropriate order in $x^2$, with the
 form of the $SU(2)$ currents shown in Eq. (\protect\ref{eq:jLcurrent}). The form appears different because 
 of the difference between $\sin\theta$, as defined in Eq. (\protect\ref{eq:deftheta}), and the ``on-shell" definition
 of $\sin\theta_W$.}
\begin{align}
g_Z & = \frac{e}{\sin\theta_W \cos\theta_W}\left[
\left(1-\frac{f^2_2}{f^2_1+f^2_2}\left\{
\epsilon^\dagger_L \epsilon_L-\frac{x^2 f^2_1}{f^2_1+f^2_2}\right\}\right)T^f_3 - Q^f \sin^2 \theta_W \right]~,
\label{eq:gZonshell}\\
g_W & = \frac{e}{\sin\theta_W}\left[
1-\frac{f^2_2}{f^2_1+f^2_2}\left\{
\epsilon^\dagger_L \epsilon_L-\frac{x^2 f^2_1}{f^2_1+f^2_2}\right\}
\right]~, \label{eq:gWonshell}
\end{align}
where $T^f_3$ and $Q^f$ are the isospin and charge of fermion species, the couplings are understood to
be matrices in flavor space, and these expressions hold up to corrections of ${\cal O}(x^4,x^2\epsilon^\dagger_L\epsilon_L)$. 
Note that, as advertised, if $\epsilon^\dagger_L \epsilon_L = (\varepsilon^{ideal}_L)^2 \cdot {\cal I} + {\cal O}(x^4)$, 
the three-site and standard-model predictions agree at tree-level up to this order. Furthermore, the deviation
of each coupling from their standard model value is proportional to
\begin{equation}
\delta g_{W,Z} \propto \frac{f^2_2}{f^2_1+f^2_2}\left\{
\epsilon^\dagger_L \epsilon_L -\frac{x^2 f^2_1}{f^2_1+f^2_2}\right\}
= \left(\frac{M_W}{M_{W'}}\right)^2 \frac{\left(\epsilon^\dagger_L \epsilon_L - (\varepsilon^{ideal}_L)^2 \cdot {\cal I}\right)}{(\varepsilon^{ideal}_L)^2}
\equiv  \left(\frac{M_W}{M_{W'}}\right)^2\frac{\Delta \epsilon^\dagger_L \epsilon_L}{(\varepsilon^{ideal}_L)^2}~,
\end{equation}
where we express the deviation in the delocalization from ideal as a fraction of
$(\varepsilon^{ideal}_L)^2$ and have used Eqs. (\ref{eq:GF}) and (\ref{eq:Wmasses}) to derive the last expression, 

This form of the three-site couplings allows comparison with the LEPEWWG \cite{:2005ema} extraction of
the (flavor-diagonal) fermion couplings to the $Z$-boson, which (in our notation) assumes the form
\begin{equation}
g^{\bar{f}f}_Z \equiv  \frac{e}{\sin\theta_W \cos\theta_W} \cdot \sqrt{\rho_f}\, (T^f_3 - Q^f \,\sin^2\theta^{eff}_W)~,
\label{eq:rhoodef}
\end{equation}
where the partial widths and asymmetries for any fermion species are
recast as measurements $\rho_f$ and $\sin^2 \theta^f_{eff}$. 
In the three-site model at tree-level, therefore, we find
\begin{align}
\sqrt{\rho^{3-site}_f} & = 1- \left(\frac{M_W}{M_{W'}}\right)^2\frac{\left[\Delta \epsilon^\dagger_L \epsilon_L\right]_{\bar{f}f}}{(\varepsilon^{ideal}_L)^2}~,
\label{eq:rhofeff}\\
\sin^2\theta^{3-site}_f & = \sin^2\theta_W \left(1+ \left(\frac{M_W}{M_{W'}}\right)^2\frac{\left[\Delta \epsilon^\dagger_L \epsilon_L\right]_{\bar{f}f}}{(\varepsilon^{ideal}_L)^2}\right)~,
\label{eq:sinsqthetaeff}
\end{align}
where $\left[\Delta \epsilon^\dagger_L \epsilon_L\right]_{\bar{f}f}$ denotes the appropriate diagonal
element of the matrix measuring the deviation of $\epsilon^\dagger_L \epsilon_L$ from ideal.

Finally assuming, for the moment, that $\epsilon^\dagger_L \epsilon_L$ is flavor-universal (proportional to the
identity),
we may use the techniques of \cite{Chivukula:2004pk} to compute the value of $\alpha S$ from the $Z$-boson
couplings to the $T_3$ and $Y$ currents
\begin{equation}
g_{3Z} \cdot g_{YZ} = -e^2 \left(1 + \frac{\alpha S}{4 \sin^2\theta \cos^2\theta}\right)~.
\end{equation}
Applying this to the expression in Eq. (\ref{eq:gZonshell}), we find
\begin{equation}
\alpha S  = -4\sin^2\theta_W\,\left(\frac{M_W}{M_{W'}}\right)^2\frac{\left[\Delta \epsilon^\dagger_L \epsilon_L\right]_{\bar{f}f}}{(\varepsilon^{ideal}_L)^2}~. \label{eq:S} 
\end{equation}

When we study the flavor structure of the quark and lepton sectors, we expect the left-handed delocalization parameter for each flavor to have a value close to $(\varepsilon_L^{ideal})^2$, and we now investigate how large a deviation is allowed by experimental data.

\subsection{The Lepton Sector}

We now consider specific experimental constraints on the lepton flavor structure in the three-site model.  By analogy with the effective Lagrangian for the quark sector (\ref{eq:Efflag}), that for the lepton sector of the three-site model may be written, defining $\ell_L \equiv (\nu,\ell^\pm)_L$, as
\begin{align}
{\cal L}_{eff} & = \bar{\ell}_{L} i \fsl{D} \ell_{L} + \bar{\ell}_{R} i \fsl{D} \ell_{R} 
-\left[ 
\bar{\ell}_{L} \epsilon_{L\ell}  \Sigma_1 \Sigma_2 \underaccent{\tilde}{\mathsf M}_{\ell} \epsilon_{R\ell}
\begin{pmatrix}
0\\
\ell_R
\end{pmatrix}
+h.c.\right] \label{eq:Efflaglep}\\
& + \bar{\ell}_{L}\epsilon_{L\ell} \left[\gamma^\mu \Sigma_1(i D_\mu \Sigma^\dagger_1)\right]
 \epsilon^\dagger_{L\ell} \ell_{L}
 + \bar{\ell}_{R} 
\epsilon_{R\ell} \left[\gamma^\mu \Sigma^\dagger_2 (iD_\mu \Sigma_2)\right]
\epsilon^\dagger_{R\ell}
\ell_{R}~,
\nonumber
\end{align}
where $\epsilon_{L\ell}$ and $\epsilon_{R\ell}$ are defined in parallel with Eqs. (\ref{eq:epsilonL}, \ref{eq:epsilonRd}).  We
use a basis where the charged lepton mass matrix
\begin{equation}
{\cal M}_\ell = \epsilon_{L\ell} \underaccent{\tilde}{\mathsf M}_{\ell} \epsilon^\dagger_{R\ell}
\end{equation}
in Eq. (\ref{eq:Efflaglep}) is diagonal, and we ignore neutrino masses.
We will focus on bounding the elements of the matrix
\begin{equation}
\eta_\ell \equiv \epsilon_{L\ell} \epsilon_{L\ell}^\dagger \,,
\label{eq:etaell}
\end{equation}
which can induce flavor-dependent $Z$ and $Z'$ couplings to the charged
leptons.  As discussed above and in \cite{SekharChivukula:2006cg}, we expect the diagonal elements of this
matrix to have values close to $(\varepsilon^{ideal}_L)^2$ so as to eliminate
contributions to $\alpha S$.

\subsubsection{Bounds on the diagonal elements of $\eta_\ell$ }
\label{subsec:diaglep}

The LEPEWWG analysis \cite{:2005ema} of $Z$ boson couplings to charged leptons constrains the diagonal elements of $\eta_\ell$.
First, under the assumption of lepton universality, we may bound the amount by which the (presumed identical) diagonal elements $\eta_{\ell ii}$ may differ from $(\varepsilon_L^{ideal})^2$.  As mentioned above, the LEPEWWG  defines a factor $\rho_f$ to accommodate the possibility that physics beyond the standard model shifts the magnitude of the $Z$ boson's coupling to the $T_3$ charge of fermion $f$ (see Eq. (\ref{eq:rhoodef})).  Under the assumption of charged lepton universality, they obtain the experimental limit $\rho_\ell = 1.0050 \pm 0.0010$, and give the standard model prediction as 1.00509$^{+0.00067}_{-0.00081}$.
Because the deviation of $\eta_{\ell ii}$ from the ideal delocalization value is proportional to the departure of $\rho_\ell$ from its value in the standard model (\ref{eq:rhofeff}),  the LEPEWWG bound on $\rho_\ell$ implies the following 90\% CL bound:
\begin{equation}
-0.036\, (\varepsilon^{ideal}_L)^2\left(\frac{M_{W'}}{400\ {\rm GeV}}\right)^2 < \eta_{\ell ii} - (\varepsilon^{ideal}_L)^2 < 0.034\, (\varepsilon^{ideal}_L)^2\left(\frac{M_{W'}}{400\ {\rm GeV}}\right)^2 
\label{eq:electron}
\end{equation}
Quantitatively similar results follow from the LEPEWWG direct experimental limit on $\sin^2\theta^{eff}_{\ell}$ and from measurements of the leptonic asymmetry ${\cal A}_e$.  We conclude that, in the case of lepton universality, the diagonal elements of $\eta_\ell$ must be within a few percent of $(\varepsilon^{ideal}_L)^2$.

Second, we may bound the degree to which the different $\eta_{\ell ii}$ may differ from one another. The LEPEWWG  has obtained the following bounds on the relative rates at which the $Z$ decays to different flavors of charged leptons  \cite{:2005ema} :
\begin{equation}
\frac{\Gamma(Z \to \mu^+\mu^-)}{\Gamma(Z \to e^+e^-)} \equiv \frac{\Gamma_\mu}{\Gamma_e} =  1.0009 \pm 0.0028\,, \quad \frac{\Gamma(Z \to \tau^+\tau^-)}{\Gamma(Z \to e^+e^-)} \equiv \frac{\Gamma_\tau}{\Gamma_e}= 1.0019 \pm 0.0032 
\end{equation}
and notes that the expected standard model values of these ratios are, respectively, 1.000 and 0.9977.   We find that these ratios are directly related to the differences between the various diagonal elements of $\eta_\ell$; for muons we have (defining $s_\theta \equiv \sin\theta$)
\begin{equation}
\frac{\delta(\Gamma_\mu/\Gamma_e)}{\Gamma_\mu/\Gamma_e} = \frac{\delta\Gamma_\mu}{\Gamma_\mu} - \frac{\delta\Gamma_e}{\Gamma_e} = \frac{(-\frac12 + s^2_\theta)\left(\frac{f^2_2}{f^2_1+f^2_2}\right)}{(-\frac12 + s^2_\theta)^2 + (s^2_\theta)^2} (\eta_{\ell 22} - \eta_{\ell 11})
\end{equation}
and a similar expression holds for taus.  The LEPEWWG limits on the ratios of partial widths thus yield (at 90\% CL)
\begin{eqnarray}
-0.063 \,(\varepsilon^{ideal}_L)^2\left(\frac{M_{W'}}{400\ {\rm GeV}}\right)^2 &<& \eta_{\ell 22} - \eta_{\ell 11} <  0.043\, (\varepsilon^{ideal}_L)^2\left(\frac{M_{W'}}{400\ {\rm GeV}}\right)^2 \,, \label{eq:mue} \\
-0.11 \,(\varepsilon^{ideal}_L)^2\left(\frac{M_{W'}}{400\ {\rm GeV}}\right)^2 &<& \eta_{\ell 33} - \eta_{\ell 11} <  0.012\, (\varepsilon^{ideal}_L)^2\left(\frac{M_{W'}}{400\ {\rm GeV}}\right)^2\,. \label{eq:taue}
\end{eqnarray}

Using the bounds on the flavor-universal lepton results  as indicative of the allowed deviation in the electron couplings
and combining the uncertainties in in Eqs. (\ref{eq:mue}) and (\ref{eq:taue}) in quadrature with those in Eq. (\ref{eq:electron}),  we find that the bounds:
\begin{eqnarray}
-0.075 \,(\varepsilon^{ideal}_L)^2\left(\frac{M_{W'}}{400\ {\rm GeV}}\right)^2 &<& \eta_{\ell 22} - (\varepsilon^{ideal}_L)^2 <  0.053\, (\varepsilon^{ideal}_L)^2\left(\frac{M_{W'}}{400\ {\rm GeV}}\right)^2 \,, \\
-0.12 \,(\varepsilon^{ideal}_L)^2\left(\frac{M_{W'}}{400\ {\rm GeV}}\right)^2 &<& \eta_{\ell 33} - (\varepsilon^{ideal}_L)^2 <  0.020\, (\varepsilon^{ideal}_L)^2\left(\frac{M_{W'}}{400\ {\rm GeV}}\right)^2\,. 
\end{eqnarray}
Hence, even without an {\it a priori} assumption of lepton universality, the diagonal elements of $\eta_\ell$ are constrained by the data to nearly equal one another.  


\subsubsection{Bounds on the off-diagonal elements of $\eta_\ell$}
\label{subsec:offdiaglep}

We now consider the bounds on the off-diagonal elements $\eta_{\ell ij}$ from lepton-flavor-violating processes.  
These arise from flavor-changing left-handed neutral-boson couplings contained in the second line of Eq. (\ref{eq:Efflaglep}).
Having diagonalized the charged-lepton mass matrix ${\cal M}_\ell$, the Hermitian flavor
matrix $\eta_\ell$ in Eq. (\ref{eq:etaell}) is fixed\footnote{In particular, the matrix $\eta$ does not change under $SU(3)_{LD} \times SU(3)_{RD}$
transformations.}  and, in general, contains off-diagonal elements. 
In unitary gauge, the gauge-operator in Eq. (\ref{eq:Efflaglep}) becomes
\begin{equation}
\Sigma^\dagger_1(D_\mu \Sigma_1)  \to( g W^a_{0\mu} -  \tilde{g}\, W^a_{1\mu}) \frac{\sigma^a}{2}~,
\end{equation}
where $g_{0,1}$ and $W_{0,1}$ are the gauge-eigenstate $SU(2)_0 \times SU(2)_1$ fields in the three-site model
in Fig. \ref{fig:one}. We may re-write the combination of neutral gauge-eigenstate fields
into mass-eigenstate fields using Eqs. (\ref{eq:Zeigenstate}, \ref{eq:Zpeigenstate})
to  find
\begin{equation}
g W^3_{0\mu} - \tilde{g}\, W^3_{1\mu} = \frac{e}{s_\theta c_\theta}\left(\frac{f^2_2}{f^2_1+f^2_2}\right) Z_\mu - \tilde{g}\, {Z'}_\mu~,
\label{eq:FCNCcombination}
\end{equation}
up to corrections of ${\cal O}(x^2)$. Note that the combination
$gW_0 - \tilde{g} W_1$ is orthogonal
to the photon; therefore, as must be true by charge conservation,
there are no flavor-changing electromagnetic couplings.

The flavor-dependent left-handed neutral-boson couplings of the leptons are, then, given by
\begin{equation}
{\cal L}_{FCNC} = \pm\frac{1}{2} \cdot \left(\frac{e}{s_\theta c_\theta} \left(\frac{f^2_2}{f^2_1+f^2_2}\right) Z_\mu - \tilde{g}{Z'}_\mu\right)
\cdot \eta_{\ell ij}\, \bar{\ell}^0_{iL} \gamma^\mu \ell^0_{jL}~,
\label{eq:Z-FCNC}
\end{equation}
Due to suppression proportional to lepton masses, the right-handed
flavor-dependent couplings are expected to be small.  In contrast to the case of meson-mixing (considered below), in the lepton sector we are interested in low-energy processes arising from only one insertion of the flavor-dependent operators. 
Hence, only the $Z^\mu$ couplings
in Eq. (\ref{eq:Z-FCNC}) contribute: the $Z'$ couplings to light fermions are suppressed.
At low energies, the flavor-dependent $Z$-boson couplings give rise
to the four-fermion operators
\begin{align}
{\cal L}_{FF}  = & \pm \frac{e^2}{2 s^2_\theta c^2_\theta M^2_Z}
\cdot \left(\frac{f^2_2}{f^2_1+f^2_2}\right) \eta_{\ell ij}\, \bar{\ell}^0_{iL} \gamma_\mu \ell^0_{jL}
\cdot J^\mu_Z + h.c. \\
= & 2\sqrt{2} G_F \cdot \eta_{\ell ij} \left(\frac{f^2_2}{f^2_1+f^2_2}\right) \, \bar{\ell}^0_{iL} \gamma_\mu \ell^0_{jL}
\cdot J^\mu_Z + h.c ~,
\label{eq:FF-FCNC}
\end{align}
where $J^\mu_Z= J^\mu_3 - Q^\mu \sin^2\theta $ is the usual 
current to which the $Z$-boson couples.

We begin with limits arising from searches for the decay $\mu \to 3e$,
where $BR(\mu^-\to e^-e^+e^-)<1.0 \times 10^{-12}$ at 90\% CL \cite{Amsler:2008zzb}.
This is easy to scale from ordinary muon decay, where the interaction
\begin{equation}
{\cal L}_{\mu-decay} = 2\sqrt{2} G_F 
(\bar{\mu}_L\gamma^\mu \nu_{L\mu})
(\bar{\nu}_{Le}\gamma^\mu e_L)
\end{equation}
yields the width
\begin{equation}
\Gamma(\mu\to e \nu_\mu  \bar{\nu}_e) =
\frac{G^2_F m^5_\mu}{192 \pi^3}~.
\end{equation}
Hence, since $BR(\mu\to e\nu_\mu \bar{\nu}_e)\simeq 100\%$, 
from Eq. (\ref{eq:FF-FCNC}) we 
find\footnote{Here the factor of ${\scriptstyle \frac{1}{2}}$ accounts for the identical particles in the $\mu \to 3e$ final
state.}
\begin{equation}
\frac{BR(\mu \to 3e)}{BR(\mu\to e\nu_\mu \bar{\nu}_e)}
\approx
\frac{1}{2} \cdot \left[\eta_{\ell 1 2} \left(\frac{f^2_2}{f^2_1+f^2_2}\right) \left(-\,\frac{1}{2} + \sin^2\theta\right)\right]^2
< 1.0 \times 10^{-12}~.
\end{equation}
This yields the bound
\begin{equation}
|\eta_{\ell 12}|  < 1.05 \times 10^{-5}\left(\frac{f^2_1+f^2_2}{2 f^2_2}\right) \ \  (90\%\ {\rm CL}) \simeq 1.3 \times 10^{-4} (\varepsilon^{ideal}_L)^2\left(\frac{M_{W'}}{400\ {\rm GeV}}\right)^2~.
\end{equation}
A quantitatively similar bound on this matrix element is found from data on $\mu\, Pb \to e\, Pb$ conversion.

By similar means, starting from the bound $BR(\tau \to e \mu \mu)<2.3 \times 10^{-8}$ at 90\% CL, we find
\begin{equation}
\frac{BR(\tau\to e\mu\mu)}{BR(\tau\to e \nu_\tau \bar{\nu}_e)} =  
\left[\eta_{\ell 1 3}\left(\frac{f^2_2}{f^2_1+f^2_2}\right)\left(-\,\frac{1}{2} + \sin^2\theta\right)\right]^2
< \frac{2.3 \times 10^{-8}}{BR(\tau\to e \nu_\tau \bar{\nu}_e)} ~.
\end{equation}
Using the fact that $BR(\tau\to e \nu_\tau \bar{\nu}_e) \simeq 18\%$, we then obtain
\begin{equation}
|\eta_{\ell 1 3} |  < 2.7 \times 10^{-3}\left(\frac{f^2_1+f^2_2}{2 f^2_2}\right) \ \ (90\%\ {\rm CL}) \simeq 3.4 \times 10^{-2} (\varepsilon^{ideal}_L)^2\left(\frac{M_{W'}}{400\ {\rm GeV}}\right)^2~.
\end{equation}
And, {\it mutatis, mutandis},  the bound $BR(\tau\to\mu e e)< 2.7 \times 10^{-8}$ at 90\% CL yields
\begin{equation}
|\eta_{\ell 2 3} |  < 2.9 \times 10^{-3}\left(\frac{f^2_1+f^2_2}{2 f^2_2}\right) \ \ (90\%\ {\rm CL}) \simeq 3.6 \times 10^{-2} (\varepsilon^{ideal}_L)^2\left(\frac{M_{W'}}{400\ {\rm GeV}}\right)^2~.
\end{equation}

\subsubsection{Lepton Summary}
\label{subsec:summlep}

Combining the 90\% CL bounds on the lepton flavor structure, therefore  we find that the deviations 
in the elements of the matrix $\eta_\ell$ 
from ideal are bounded by:
\begin{equation}
|\eta_\ell - (\varepsilon^{ideal}_L)^2\cdot {\cal I}| \laem (\varepsilon_L^{ideal})^2 \left(\frac{M_{W'}}{400\ {\rm GeV}}\right)^2 
\left(
\begin{array}{ccc}
0.036 & 0.00013 &  0.034 \\ 
0.00013 & 0.075 &  0.036 \\ 
 0.034 &  0.036 & 0.12\end{array}\right)
\end{equation}
and $\eta_\ell$ is therefore essentially constrained to be proportional to the identity, with diagonal elements equal to $(\varepsilon^{ideal}_L)^2$.

\subsection{The Quark Sector}
\label{sec:qfcnc}

In this section we study the left-handed quark delocalization matrix $\eta = (\epsilon^\dagger_L \epsilon_L)$, introduced in 
Eq. (\ref{eq:defineeta}).  Using data on flavor-changing neutral currents and $Z$ decays to heavy quarks, we set bounds on the degree to which $\eta$ can deviate from $(\varepsilon^{ideal}_L)^2 \cdot {\cal I}$.

\subsubsection{Flavor-Changing Neutral Currents}

We begin with the most severely constrainted interactions: 
the flavor-changing left-handed neutral-boson couplings contained in the second line of equation (\ref{eq:Efflag}).  Retracing the analysis of lepton-flavor-violation above shows that, at low energies,  $Z$ and $Z'$ exchange (see Eq. (\ref{eq:FCNCcombination})) between quarks gives rise to four-fermion operators of the form
\begin{equation}
{\cal L}_{L-FCNC} \to \pm\,\frac{1}{2!}\cdot\left(\frac{1}{2}\right)^2 \cdot \eta_{ij} \eta_{k\ell} 
\left[ \frac{e^2}{ s^2_\theta c^2_\theta}\left(\frac{f^2_2}{f^2_1+f^2_2}\right)^2 \frac{1}{M^2_Z} + \frac{\tilde{g}^2}{M^2_{Z'}}\right]
(\bar{q}^i_L \gamma^\mu q^j_L) (\bar{q}^k_L \gamma_\mu q^\ell_L)~,
\label{eq:quarkfcnceff}
\end{equation}
here the first factor ($1/2!$) accounts for the two identical currents and the next ($(1/2)^2$) accounts
for the $T_3$ charges of the external fermions. Using the masses of Eq. (\ref{eq:Wmasses}) and the
relation in Eq. (\ref{eq:GF}), we find the term in square brackets is approximately $4/f_1^2$ so that
\begin{equation}
{\cal L}_{L-FCNC} \to \pm\, \frac{\eta_{ij} \eta_{k\ell} }{2 f^2_1} 
(\bar{q}^i_L \gamma^\mu q^j_L) (\bar{q}^k_L \gamma_\mu q^\ell_L)~.
\label{eq:fourfermi-fcnc}
\end{equation}

Ref. \cite{Bona:2007vi} has derived constraints on a variety of $\Delta F = 2$ four-fermion operators that cause neutral meson mixing.  We will start with their limits on the coefficients ($C^1_j$) of the operators responsible for mixing in the Kaon, $B_d$, and $B_s$ systems:
\begin{eqnarray}
C^1_K (\bar{s}_L \gamma^\mu d_L) (\bar{s}_L \gamma_\mu d_L) \qquad\qquad C^1_{B_d} (\bar{b}_L \gamma^\mu d_L) (\bar{b}_L \gamma_\mu d_L) \qquad\qquad
C^1_{B_s} (\bar{b}_L \gamma^\mu s_L) (\bar{b}_L \gamma_\mu s_L) \, . \label{eq:cdeffs}
\end{eqnarray}
The numerical values of the limits they obtain in the down-quark sector in the $C^1_j$ correspond, in the notation of 
 Eq. (\ref{eq:fourfermi-fcnc}), to the constraints
\begin{align}
- (4.82 \times 10^{-4})^2< \Re(\eta_{sd})^2 \left(\frac{2v^2}{f^2_1}\right)  & < (4.82 \times 10^{-4})^2 \label{eq:fcnc1}\\
-(3.26 \times 10^{-5})^2 <  \Im (\eta_{sd})^2 \left(\frac{2v^2}{f^2_1}\right) & < (2.60 \times 10^{-5})^2 \label{eq:fcnc2}\\
|\eta_{bd}|^2 \left(\frac{2v^2}{f^2_1}\right) & < (2.3 \times 10^{-3})^2  \label{eq:fcnc3}\\
|\eta_{bs}|^2 \left(\frac{2v^2}{f^2_1}\right) & < (1.63 \times 10^{-2})^2 \label{eq:fcnc4}
\end{align}
or, in a more convenient notation, to
\begin{align}
|\eta_{ds}| & < 4.8 \times 10^{-4} \left(\frac{f_1}{\sqrt{2}v}\right)= 0.0060 \, (\varepsilon^{ideal}_L)^2\left(\frac{M_{W'}}{400\ {\rm GeV}}\right)^2\left(\frac{\sqrt{2}v}{f_1}\right)\label{eq:fcnci1}\\
|\eta_{bd}| & < 2.3 \times 10^{-3}  \left(\frac{f_1}{\sqrt{2}v}\right) = 0.0285\, (\varepsilon^{ideal}_L)^2\left(\frac{M_{W'}}{400\ {\rm GeV}}\right)^2\left(\frac{\sqrt{2}v}{f_1}\right)\label{eq:fcnci2}\\
|\eta_{bs}| & < 1.63 \times 10^{-2}\left(\frac{f_1}{\sqrt{2}v}\right)  = 0.202\, \, (\varepsilon^{ideal}_L)^2\left(\frac{M_{W'}}{400\ {\rm GeV}}\right)^2\left(\frac{\sqrt{2}v}{f_1}\right)\label{eq:fcnci3}
\end{align}
In the three-site model, we expect the eigenvalues of the matrix
$\eta$ to be of order $(\varepsilon^{ideal}_L)^2$.
Hence, with the possible exception of $\eta_{bs}$, the data requires that the matrix 
$\eta $ be nearly {\it diagonal} in the down-quark mass-eigenstate basis.  

At this point, recalling that ${\cal M}_u = V^\dagger_{CKM} \Delta_u$, we also note that there is a low-energy operator that can cause $D$-meson mixing.  This is
\begin{equation}
 C^1_D\, (\bar{c}_L \gamma^\mu u_L) (\bar{c}_L \gamma_\mu u_L)~,
 \end{equation}
with
\begin{equation}
C^1_D = \pm \frac{1}{2f^2_1}
\left( V_{ud}\, \eta_{11} V^*_{cd} + V_{us}\, \eta_{22} V^*_{cs} + V_{ub}\, \eta_{33} V^*_{cb}\right)^2~,
\end{equation}
where the $V_{ij}$ are the elements of the CKM matrix.
The authors of \cite{Bona:2007vi} report a limit 
\begin{equation}
\vert C^1_D  \vert < 7.2 \times 10^{-13}\, \rm{GeV}^{-2} \,,
\end{equation}
from which we conclude
\begin{equation}
|V_{ud}\, \eta_{11} V^*_{cd} + V_{us}\, \eta_{22} V^*_{cs} + V_{ub}\, \eta_{33} V^*_{cb}|^2 < (4.17 \times 10^{-4})^2 \left(\frac{f^2_1} {2v^2}\right)~.
\label{eq:etaD-bound}
\end{equation}
Now, the product of CKM elements appearing in the third term $|V_{ub} V^*_{cb}| \simeq {\cal O}(10^{-4})$ is much smaller than those in the other two terms
$V_{ud} V^*_{cd} \approx -V_{us} V^*_{sc} \approx .16$. Therefore,  barring
a very large difference among the diagonal entries of $\eta$,
we may neglect the $\eta_{33}$ term in Eq. (\ref{eq:etaD-bound}) and find
\begin{equation}
|\eta_{11}-\eta_{22}| \le 2.61 \times 10^{-3}   \left(\frac{f_1}{\sqrt{2}v}\right) = 0.0323 \, (\varepsilon^{ideal}_L)^2\left(\frac{M_{W'}}{400\ {\rm GeV}}\right)^2\left(\frac{\sqrt{2}v}{f_1}\right)
\end{equation}
Since we anticipate that each of the $\eta_{ii}$ is of order $(\varepsilon_L^{ideal})^2$,  we conclude that $\eta_{11} \approx \eta_{22}$.

This result is 
consistent with precision electroweak data: $\eta_{11}$ and $\eta_{22}$ respectively, determine
the delocalization of the first- and second-generation left-handed quarks. Their having different values is disfavored because that would change the relative rates at which the $Z$ decays to up vs. charm or down vs. strange quarks. Similarly, $\eta_{33}$ controls the delocalization of $b_L$ -- and, as discussed below, data on $R_b$ and $R_c$ constrains how much this can differ from $\eta_{11,22}$.

These are the strongest limits available from flavor-changing processes in the quark sector. Bounds on flavor-changing decays in the third generation up-quark sector are rather weak: current limits imply only that $Br(t \to c Z) < 3.7\%$ \protect\cite{Amsler:2008zzb}, which provides no new information on the elements of $\eta$.  While the $B_s$ or $D^0$ systems are, respectively, the most promising for eventual limits on right-handed FCNC's in the down and up sectors, no limits presently exist.

\subsubsection{$Z$-Pole Constraints on $R_b$ and $R_c$}
\label{subsubsec:eta2233rbrc}

The LEPEWWG  has obtained bounds on the relative rates at which the $Z$ decays to heavy quarks, as compared with decays to all hadrons  \cite{:2005ema} :
\begin{eqnarray}
\frac{\Gamma(Z \to b\bar{b})}{\Gamma(Z \to hadrons)} &\equiv& R_b =  0.21629 \pm 0.00066\\
\frac{\Gamma(Z \to c\bar{c})}{\Gamma(Z \to hadrons)} &\equiv& R_c= 0.1721 \pm 0.0030 \, ,
\end{eqnarray}
and gives the, respective, standard model predictions for these quantities as $0.21583^{+0.00033}_{-0.00045}$ and $0.17225^{+0.00016}_{-0.00012}$.  These ratios are useful to work with because QCD corrections, manifesting as dependence on the value of $\alpha_s$, should largely cancel.\footnote{In principle, one could try to extract limits on $\eta_{33}$ from the product $R_{b} R_{\ell}$, because the fractional change would depend only on $\eta_{33}$ and $\eta_{\ell ii}$, and the latter is already tightly constrained to have the value $(\varepsilon_L^{ideal})^2$. However, the usefulness of this approach is limited by the fact that the standard model prediction of $R_\ell$ is subject to significant uncertainty through its dependence on $\alpha_s$.}

Because the data from D-meson mixing has already established that $\eta_{22} = \eta_{11}$, both $R_b$ and $R_c$ may be written as linear combinations of just the two diagonal matrix elements $\eta_{33}$ and $\eta_{22}$:
\begin{eqnarray}
\frac{\delta(R_b)}{R_b} &=& \frac{\delta\Gamma_b}{\Gamma_b} - \frac{\delta\Gamma_{hadr.}}{\Gamma_{hadr.}} = \left(-0.8924\, ( \eta_{33} - (\varepsilon^{ideal}_L)^2) + 0.0910\, ( \eta_{22} - (\varepsilon^{ideal}_L)^2)\right)\left[\frac{2f^2_2}{f^1_1+f^2_2}\right] \label{eq:firs-coupled} \\
\frac{\delta(R_c)}{R_c} &=& \frac{\delta\Gamma_c}{\Gamma_c} - \frac{\delta\Gamma_{hadr.}}{\Gamma_{hadr.}} = \left(0.2512\, (\eta_{33} - (\varepsilon^{ideal}_L)^2) + 1.297\, (\eta_{22} - (\varepsilon^{ideal}_L)^2)\right) \left[\frac{2f^2_2}{f^1_1+f^2_2}\right]\,, \label{eq:sec-coupled}
\end{eqnarray}
Solving the coupled equations for the two $\eta_{ii} - (\varepsilon^{ideal}_L)^2$ yields the limits
\begin{eqnarray}
-0.093 \,(\varepsilon^{ideal}_L)^2\left(\frac{M_{W'}}{400\ {\rm GeV}}\right)^2 &<& \eta_{33} - (\varepsilon^{ideal}_L)^2 <  0.020\, (\varepsilon^{ideal}_L)^2\left(\frac{M_{W'}}{400\ {\rm GeV}}\right)^2 \,, \\
-0.30 \,(\varepsilon^{ideal}_L)^2\left(\frac{M_{W'}}{400\ {\rm GeV}}\right)^2 &<& \eta_{22} - (\varepsilon^{ideal}_L)^2 <  0.30\, (\varepsilon^{ideal}_L)^2\left(\frac{M_{W'}}{400\ {\rm GeV}}\right)^2\,.
\end{eqnarray}
while rotating (\ref{eq:firs-coupled}) and (\ref{eq:sec-coupled}) into the $\eta_{33} \pm \eta_{22}$ basis says, equivalently:
\begin{equation}
-0.48\,(\varepsilon^{ideal}_L)^2\left(\frac{M_{W'}}{400\ {\rm GeV}}\right)^2 < \eta_{33} - \eta_{22}  <  0.41\, (\varepsilon^{ideal}_L)^2\left(\frac{M_{W'}}{400\ {\rm GeV}}\right)^2\,.
\end{equation}

We conclude that $\eta_{33}$ is constrained at 90\% CL to lie within a few percent of the ideal delocalization value, while $\eta_{22}$ (and $\eta_{11}$) must lie within about 30\% of the ideal delocalization value and within about 45\% of $\eta_{33}$. The limit on $\eta_{22}$ is consistent with what the LEPEWWG data on $\rho_{c}$ implies; the limit on $\eta_{33}$ surpasses that obtained from $\rho_b$.

\subsubsection{Summary}
\label{subsec:summquark}

Combining the 90\% CL bounds for the $\eta_{ii}$ obtained in this section, we find that  deviations in the elements of the matrix $\eta$ from ideal 
delocalization are bounded by:
\begin{equation}
\vert \eta - (\varepsilon^{ideal}_L)^2\cdot {\cal I} \vert  \laem 
(\varepsilon_L^{ideal})^2 \left(\frac{M_{W'}}{400\ {\rm GeV}}\right)^2 \left(\begin{array}{ccc}0.30 &  0.0060 \frac{\sqrt{2}v}{f_1} &  0.0285 \frac{\sqrt{2}v}{f_1} \\ 
0.0060 \frac{\sqrt{2}v}{f_1} & 0.30 & 0.202 \frac{\sqrt{2}v}{f_1} \\ 0.0285 \frac{\sqrt{2}v}{f_1}&  0.202 \frac{\sqrt{2}v}{f_1} & 0.09\end{array}\right)\ ,
\end{equation}
subject to the constraints on $\eta_{22} - \eta_{11}$ and $\eta_{33} - \eta_{22}$ noted above.   The factors of $ \sqrt{2} v / f_1$ in the off-diagonal elements reflect the fact that those bounds arise from joint $Z$ and $Z'$ contributions to $\Delta F = 2$ meson mixing processes;  the constraints on the diagonal elements, like all the elements of $\eta_\ell$, come from decay processes involving only $Z$ couplings.  We conclude that the flavor matrix $\eta$ for quarks must be nearly proportional to the identity  matrix.

\end{document}